\documentclass[symmetry,review,accept,pdftex,moreauthors]{Definitions/mdpi} 

\DeclareMathOperator{\Tr}{Tr}

\firstpage{1} 
\pubvolume{15}
\issuenum{1}
\articlenumber{1497}
\pubyear{2023}
\copyrightyear{2023}
\externaleditor{Academic Editor: Michal Hnatič and Juha Honkonen}
\datereceived{29 June 2023} 
\daterevised{21 July 2023} 
\dateaccepted{26 July 2023} 
\datepublished{28 July 2023} 
\hreflink{https://doi.org/10.3390/sym15081497} 
\doinum{10.3390/sym15081497}
\pdfoutput=1 

\Title{Perturbative Asymptotic Safety and Its\linebreak  Phenomenological Applications}

\TitleCitation{Perturbative Asymptotic Safety and Its Phenomenological Applications}

\Author{Alexander Bednyakov *$^{,\dagger}$\orcidA{} and Alfiia Mukhaeva *$^{,\dagger}$\orcidB{}}

 \AuthorNames{Alexander Bednyakov, Alfiia Mukhaeva}

\AuthorCitation{Bednyakov, A.;\linebreak  Mukhaeva, A.}

\address[1]{%
Joint Institute for Nuclear Research, Joliot-Curie 6, 141980 Dubna, Russia}

\corres{\hangafter=1 \hangindent=1.05em \hspace{-0.82em}Correspondence:  
bednya@jinr.ru 
 (A.B.); mukhaeva@theor.jinr.ru (A.M.)}

\firstnote{\hangafter=1 \hangindent=1.05em \hspace{-0.82em}These authors contributed equally to this work. 
}

\abstract{ Asymptotic safety is a remarkable example when fruitful ideas borrowed from statistical physics proliferate to high-energy physics. The concept of asymptotic safety is tightly connected to fixed points (FPs) of the renormalization-group (RG) flow, and generalize well-known asymptotic freedom to a scale-invariant ultraviolet completion with non-vanishing interactions. In this review, we discuss the key ideas behind asymptotic safety, a mechanism for achieving it, and the conditions it imposes on general gauge--Yukawa field theories. We also pay special attention to possible phenomenological applications and provide an overview of standard model (SM) extensions potentially exhibiting asymptotic safety. }

\begin{document}
\section{Introduction\label{sec:intro}}

Today, we know two very successful theories of nature: The standard model (SM) and Einstein gravity. Both of them are thoroughly tested in various experiments. Despite the presence of some tensions, there is not really any conclusive indication that these theories are insufficient to describe the nature at scales at which we are currently testing them. 
 However, neither of those two seems ultraviolet (UV) complete and can only be treated as effective field theories (EFT) valid at relatively low scales.

The SM, while being a formally renormalizable Quantum field theory (QFT), exhibits singularities in the far UV---Landau poles in scale-dependent couplings such as the Abelian hypercharge and that of the Higgs--Yukawa sector. It is quite interesting that the scale at which the SM itself breaks down is trans-Planckian (far above the Planck mass). Therefore, this fact opens up the possibility that the quantum gravity can provide the UV-extended or even completed standard model. %

When considering Einstein gravity as EFT, the breakdown of predictivity is directly connected to the theory's perturbative non-renormalizability. This non-renormalizability necessitates the introduction of an infinite number of counter terms to absorb arising UV divergencies, each associated with its own, a priori arbitrary, coupling constant that should be fixed from the experiment. Consequently, the theory ends up with an infinitely large number of free parameters, which ultimately undermines predictivity. %

The problem of the UV divergences and infinite number of free parameters of EFTs can be addressed in the context of Asymptotic Safety (AS).  %
The idea of AS was proposed by\linebreak  S. Weinberg \cite{Weinberg:AS} as a way of making the four-dimensional theory of gravity non-perturbatively renormalizable in the late 1970s 
 and 
is tightly connected to quantum version of\linebreak scale~invariance. 

Generically, QFTs are not scale invariant, i.e., in the presence of quantum fluctuations, the scaling symmetry is broken and features a non-trivial renormalization group (RG) flow in a theory (coupling) 
 space. 
This means that the couplings entering the QFT action become dependent on the energy or momentum scale and that effective dynamics changes as you go from scale to scale. The initial condition corresponds to the bare (microscopic) action in the UV (defined with a certain cutoff $\Lambda$), and the flow towards the infrared (IR) gives rise to a trajectory in the coupling space.

In general, the theory space is infinitely dimensional and accounts for all possible operators that are compatible with symmetries of the action 
(since operators are built from quantum fields, there is a freedom in the basis 
 choice of the latter that can be translated to the freedom in the coupling space. In what follows, we only consider essential couplings that can not be removed by field redefinitions (see, e.g, Ref.~\cite{Percacci:2007sz})).   
If the RG flow features a fixed point (FP) the scaling symmetry can be recovered at the quantum level, resulting,\linebreak  e.g., in the possibility to remove the UV cutoff ($\Lambda \to \infty$). For example, a well-known asymptotic freedom corresponds to a trivial restoration of scale symmetry in the sense that it switches off all interactions ( the so-called Gaussian FP) and, thus, removes the effect of quantum fluctuations completely. Another possibility is an interacting (or partially interacting) RG fixed point at finite values of the couplings. This latter case is utilised in the Asymptotic safety framework.

Near FPs, the operators entering the bare action can be ordered by the corresponding critical exponents, which, at the Gaussian FP, coincide with canonical dimensions of the couplings (see below). 
The (combinations of) operators with negative exponents are said to be irrelevant in the IR, since the corresponding couplings are attracted to the FP values as we decrease the scale. In this respect, we have a prediction in the IR, e.g., a fixed value, or, 
 in a more general situation, a relation between certain couplings. 

On the contrary,  positive critical exponents give rise to directions in the coupling space that are repelled from the FP along the RG flow towards IR, and, thus, can not be predicted from the FP values; tiny deviations in the bare action can have drastic consequences in the IR. These, relevant, directions span what is called a UV-critical surface, and the number of independent directions constitutes the number of free parameters of the theory that should be eventually fixed from the experiment.  Contrary to general EFT, in which the couplings of different operators are thought to be independent, in AS scenarios, physical trajectories are assumed to reside on this finite dimensional submanifold in the infinite dimensional theory space. 
As a consequence, the lower the dimensionality of the UV critical surface, the more predictive the theory is. Notably, all irrelevant couplings can deviate from the fixed point along the critical surface. 

In this respect, one overcomes the issue with infinite number of free parameters.\linebreak  The problem of possible UV singularities is also addressed in this case, since by reversing the flow towards UV (corresponding to $\Lambda \to \infty$), one reaches the FP with finite values of all the couplings.

This is the essence of fundamental asymptotic safety. One can also envisage a non-fundamental AS, for which the bare action (at finite cutoff) is chosen (slightly) off the UV critical surface of considered FP. In this case, we can not safely extrapolate $\Lambda \to \infty$ (unless we hit another FP), since the flow towards UV is repelled from the surface. However, in the IR, the couplings are attracted to the FP and we again have predictions at low scales.

While a non-perturbative determination of the RG flow is quite involved and usually based on the functional renormalization group (FRG), a remarkable progress is achieved in perturbative RG, in which the equations that drive the flow can be computed order-by-order in loop expansion around the Gaussian FP. As an example, we refer to the convenient possibility of extracting necessary equations in a general renormalizable quantum-field theory in $d=4$ dimensions via various computer codes \cite{Staub:2013tta,Thomsen:2021ncy,Litim:2020jvl} that can combine old~\cite{Machacek:1983tz,Machacek:1983fi,Machacek:1984zw,Luo:2002ti} and new~\cite{Schienbein:2018fsw,Poole:2019kcm,Bednyakov:2021qxa,Davies:2021mnc,Jack:2023zjt} results. 

In this mini-review, we mainly rely on perturbative RG and consider particle-physic implications of AS. In spite of the fact that asymptotic safety was initially proposed to make quantum gravity self-consistent, we avoid this topic as much as possible in the review. Nevertheless, let us give some important comments on AS gravity. %

At the end of the 1990s, M.~Reuter and F.~Saueressig \cite{Reuter:1996cp,Reuter:2001ag} considered a very simple gravitational Einstein--Hilbert action, that has only two operators parametrised by the dimensionless Newton constant and vacuum energy. They found two types of fixed points.\linebreak  The first one is non-interacting (Gaussian) FP. The second one is the UV interactive FP, at which both gravitational constants are non-zero. This fixed point would actually correspond to high-energy regime so other gravitational interactions may become important and spoil the FP existence. 

To address this issue, there has been a lot of activity and more elaborated calculations, which demonstrated that such a fixed point is not really an artifact of simplification. Even if we start to add more higher order operators to this action, such FP always persists, see, for example, Refs.~\cite{Lauscher:2002sq,Litim:2003vp,Codello:2006in,Machado:2007ea,Codello:2008vh,Benedetti:2009rx,Dietz:2012ic,Falls:2013bv,Falls:2014tra,Gies:2016con}, and references therein. 

However, there exist open questions, which are discussed in more detail in Ref.~\cite{Bonanno:2020bil}. Among the issues are the background and gauge-fixing dependence of the results obtained in quantum gravity. The authors of Refs.~\cite{Morris:2015oca,Becker:2014qya,Denz:2016qks} are making first steps in trying address some of these problems.
Moreover, the renormalization procedure requires higher-order\linebreak  (in curvature) operators to be added to the Einstein--Hilbert action, giving rise to potentially ghost-like instabilities. 
The following more recent Refs.~\cite{Draper:2020bop,Platania:2020knd,Bonanno:2021squ} demonstrate ways of constructing effective dynamics involving a bunch of these higher order terms, but nevertheless without any tachyonic instabilities. 
In addition, most of the computations are carried out with metrics having Euclidean signatures. 
 Thus, an understanding of how that carries over to the Lorentzian signature is another really critical open issue, see, e.g., Ref.~\cite{Manrique:2011jc}. 

One can also ask an important question regarding the influence of matter on gravity in the context of asymptotic safety. It is known that even minimal coupling to the gravity of a self-interacting scalar field $\phi$ can give rise to non-zero non-minimal interactions of the form $\xi \phi^2 R$ with curvature $R$, when quantum corrections from matter fields are taken into account (see, e.g., Ref.~\cite{Buchbinder:1984ebp}). This coupling seems to violate a strong equivalence principle but is very important for (Higgs) inflation scenarios such as that given, e.g., in Ref.~\cite{Bezrukov:2007ep}. A recent study of Ref.~\cite{Eichhorn:2020sbo} considers the issue of obtaining correct values of the slow-roll parameters within AS in the SM-like models with scalars and fermions. 
While we appreciate the importance of these kind of studies, we also refrain from touching this topic in this review and will return back to particle physics. %

In recent years, asymptotic safety has been quite extensively used when dealing with the triviality problem of the $U(1)$ gauge couplings by making the latter reach the interactive fixed point at some scale \cite{Eichhorn:2017lry}. 
Moreover, after the discovery of the Higgs boson \cite{ATLAS:2012yve,CMS:2012qbp},\linebreak  we know that the standard model can consistently be extended up to the Planck\linebreak  scale~\cite{Bezrukov:2012sa,Buttazzo:2013uya,Bezrukov:2014ina,Bednyakov:2015sca}. Subsequently, the interaction of the standard model with quantum fluctuations of gravity has also been actively studied in the framework of quantum field theory~\cite{Bond:2017wut,Barducci:2018ysr,Hiller:2020fbu}. 
Progress in studying asymptotically safe theories has also been made in the context of supersymmetric models~\cite{Bond:2017suy}, conformal windows of parameters~\cite{Bond:2017tbw}, and within the models possessing large particle multiplicities~\cite{Pelaggi:2017abg,Antipin:2018zdg,Abel:2018fls,Alanne:2019vuk,Leino:2019qwk}. 

Recently, proposals have been put forward that connect asymptotic safety with flavour physics within and beyond the SM \cite{Hiller:2019mou,Kowalska:2020gie}. Indeed, it has been demonstrated that AS models may be able to explain measurements in the flavour sector, in particular, with discrepancies with the SM predictions.  Moreover, asymptotically safe SM extensions both with and without taking into account quantum gravity effects can explain  the flavour pattern of the SM \cite{Alkofer:2020vtb,Kowalska:2020gie,Kowalska:2022ypk}.  Altogether, asymptotically safe UV completions of the SM can present strong implications for flavour physics.

This paper is organised as follows. In Section~\ref{sec:gauge_yukawa}, we introduce key ideas and notions of asymptotic safety. We consider the RG flow in a simple, yet general, gauge--Yukawa model, discuss the fixed points of the flow and enumerate different phases that can be achieved by varying the gauge group and matter-field representations in Section~\ref{sec:toy_model}. We switch to realistic SM extensions in Section~\ref{sec:portals} and review some of BSM scenarios available on the market together with their phenomenological applications. When considering models with matter coupled to gravity in Section~\ref{sec:gravity_matter_models}, we follow a pragmatic approach to gravity-induced corrections and discuss how the ideas behind AS can enhance the predictive power of New Physics (NP). Our conclusions can be found in Section~\ref{sec:conclusions}.

\section{Asymptotic Safety in Gauge--Yukawa Theories \label{sec:gauge_yukawa}}

As a starting point, we consider the space of dimensionless couplings $g_i$ that enter\linebreak  a general action of, not necessarily, 
 a renormalizable theory in $d$ space--time dimensions: 
\begin{align}
	S = \int d^d x \mu^{d-\Delta_i} g^i O_i(x)
\end{align}
with $O_i$ being a set of local operators with scaling dimensions $\Delta_i$. The scale $\mu$ denotes the RG scale. The RG flow is driven by beta functions and is described by first-order differential renormalization-group equations (RGE):
\begin{align}
\partial_t \alpha_i = \beta_i(\alpha), \qquad \alpha_i \equiv \frac{g_i^2}{16 \pi^2}, \quad t = \ln \mu,
\label{eq:rg_beta_def}
\end{align}
where for convenience, we introduce  $\alpha_i$ for every $g_i$. 
In perturbation theory, we have the following expansion
\vspace{-3pt}
\begin{align}
	\beta_i(\alpha) = \beta_i^{(1)} + \beta_i^{(2)} + \ldots 
\end{align}
with $\beta_i^{(l)}$ corresponding to the $l$-loop correction. For given initial values $\alpha_i(0)$ of the couplants, the flow towards infrared (IR) corresponds to $t\to -\infty$, while in the limit $t \to \infty$ we approach the UV region. As required by asymptotic safety, the $\beta$-functions of the theory represent a fixed point, i.e., some set of non-trivial coupling values  $\alpha^*$, for which all $\beta$-functions vanish.
This condition can be expressed as
\begin{align}
    \beta_i(\alpha)|_{\alpha=\alpha^*} %
    = 0.
    \label{eq:beta_zero}
\end{align}

An RG trajectory that ends in the UV at such a fixed point corresponds to a UV-complete theory \cite{Percacci:2011fr}, which remains meaningful at all scales. Such RG trajectories give rise to a ``fundamental'' asymptotic safety. However, it is also worth considering a  ``non-fundamental'' case arising when an FP is a saddle-point possessing both UV- and IR-attractive directions. In such a situation, it provides a UV completion only for some RG trajectories, while acting as an IR attractor for a more fundamental description \cite{Eichhorn:2018yfc}.

When looking for fixed points, we will demand the following:
\begin{itemize}
	\item[(i)] The coordinates must be physical, fulfilling $\alpha^* \geq 0$;
	\item[(ii)] Couplings must be perturbative ({for more elaborate conditions of perturbativity,\linebreak  see, e.g., Ref.~\cite{Barducci:2018ysr}}), which requires $\alpha^* \leq 1$.
\end{itemize}

The former condition reflects the fact that $\alpha_i$ is a square of $g_i$, while the latter allows one to choose weakly interacting fixed points that can potentially render the theory predictive at all scales.  In a model with some external parameters, e.g., the number of colours $N_c$ or field species $N_f$, the solution $\alpha_i = \alpha_i^*$ of \eqref{eq:beta_zero} depends on these quantities and usually exists only for values lying in particular intervals (``windows'').

In order to illustrate the instances in which a model can present such fixed points,
we now study a simple renormalizable gauge--Yukawa theory in $d=4$ dimensions containing one gauge ($\alpha_g$) and one Yukawa ($\alpha_y$) coupling. Following Refs.~\cite{Litim:2014uca,Bond:2016dvk} and related works, we use $kmn$-ordering corresponding to a $k$-loop RGE for the gauge, $m$-loop RGE for Yukawa, and $n$-loop RGE for scalar self-couplings, and consider, for simplicity, the $210$-case. 

Here, we should note about Weyl consistency conditions (WCC), which relate derivatives of beta functions \cite{Jack:2013sha}. They arise by considering a model on a curved (but fixed) background and performing Weyl rescalings of the metric. Due to the fact that two subsequent Weyl rescalings commute, it follows that $\frac{\partial\beta^i}{\partial g_j} = \frac{\partial\beta^j}{\partial g_i}$. Herein, $\beta^i=\chi^{ij}\beta_j$,\linebreak  where $\chi^{ij}$ is a metric in the space of couplings that depends on the latter. An expression for $\chi^{ij}$ for gauge--Yukawa models in the 321-approximation has been derived in~\cite{Antipin:2013pya}, while the 432-case in a general renormalizable field theory was considered in \cite{Poole:2019kcm}.  These conditions must be satisfied for the full RG flow and can be imposed on the perturbative expansion. It is worth mentioning that Ref.~\cite{Bond:2017tbw} discusses different ordering schemes for beta functions in the context of gauge--Yukawa theories (see also Section~\ref{sec:toy_model}). 

In the 210-approximation, the scalar self-interactions decouple and we can restrict ourselves to the $\beta$-functions 
\vspace{-3pt}
\begin{align}
	\beta_g & = \alpha_g^2(-B+C\alpha_g-D\alpha_y),
    \label{eq:gy_beta_g} \\
	\beta_y & = \alpha_y(E\alpha_y-F\alpha_g).
    \label{eq:gy_beta_yu}
\end{align}

Here $B, E, F$ are one-loop coefficients, while $C, D$ come from two loops. 
While $E$,\linebreak  ($F$ and $D$) are assumed to be positive (non-negative), the signs of $C$ and $B$ depend on the specific particle content and symmetries of a theory.

For a single-gauge group with $n_f$ charged Weyl ($\kappa=1/2$) or Dirac ($\kappa=1$) fermions, and $n_s$ charged scalars, we can write \cite{Gross:1973id,Jones:1974mm,Tarasov:1976ef}
\begin{align}
	B & = 2 \left[
		\frac{11}{3} C_A - \frac{4}{3} \kappa (T_f n_f) - \frac{1}{6}(T_s n_s) 
	\right], \\
	  C & = 2 \left[
			-\frac{34}{3}C_A^2 + \kappa \left(4 C_f + \frac{20}{3} C_A\right)(T_f n_f)  
			+ \left( 2 C_s + \frac{1}{3} C_A \right) (T_s n_s)
		\right].
\end{align}	

Here, $C_A$ is the second  Casimir for adjoint representation, while $C_{R}$ and $T_{R}$ refer to the quadratic Casimirs and the Dynkin index, respectively, for fermion ($R=f$)\linebreak  and scalar ($R=s$) representations.
For $SU(N)$ gauge theory with fermions in fundamental representation, we have $C_A = N_c$, $C_f = (N_c^2 - 1)/(2N_c)$, and $T_f = 1/2$. Obviously,\linebreak  for Abelian gauge groups $C_A=0$; thus, we always have $B<0$ irrespectively of matter content, while for non-Abelian theories, negative contributions from charged fermions and scalars can be compensated  by that of gauge field fluctuations. 

In Ref.~\cite{Bond:2016dvk}, Bond and Litim studied possible signs of $B$ and $C$. Utilizing the relation 
\begin{align}
    C = \frac{2}{11}
    \left[
	    2 \kappa
	    \left(
	    11 C_f + 7 C_A
    \right) (n_f T_f)
    + 2 \left( 11 C_s - C_A\right) (n_s T_s)
    - 17 C_A \cdot B
    \right]
    \label{eq:C_dep_B}
\end{align}
they demonstrated that for $B\leq0$ all the contributions are positive and render $C>0$ irrespectively of matter representations, while for $B>0$, the two-loop coefficient $C$ can be both negative and positive. 
This information is crucial when studying the behaviour of the RG flow and the possibility of asymptotic safety in the gauge--Yukawa models.

Several types of fixed points exist for the system \eqref{eq:gy_beta_g} and \eqref{eq:gy_beta_yu}. Firstly, the Gaussian FP\linebreak  is given by
\begin{align}
    \alpha_g^* = \alpha_y^* = 0,
    \label{eq:GA_fp}
\end{align}
and may present itself in different energy regimes (IR or UV).
The second option is when Equations~\eqref{eq:gy_beta_g} and \eqref{eq:gy_beta_yu} admit a fixed point for which the Yukawa $\alpha_y$ is asymptotically free\linebreak  (in the IR), but the gauge $\alpha_g$ is interacting: %
\begin{align}
    \alpha_g^*=\frac{B}{C}, \qquad \alpha_y^*=0.
    \label{eq:CBZ_fp}
\end{align}

The above solution is known as the Caswell--Banks--Zaks (BZ) FP \cite{Caswell:1974gg,Banks:1981nn}. It requires $B/C > 0$ in order to be physical and for $B/C < 1$, it can be treated in perturbation theory. 

Finally, the system develops another type of FP, where both couplings are non-vanishing. This is the gauge--Yukawa (GY) FP, which is characterised by the coordinates
\begin{align}
    \alpha_g^*=\frac{B}{C^\prime}, \qquad \alpha_y^*=\frac{F}{E} \alpha_g^* = \frac{F B}{E C'},
    \label{eq:GY_fp}
\end{align}
where  the coefficient 
\vspace{-3pt}
\begin{align}
    C^\prime = C - \frac{DF}{E} \leq C
\end{align}
can take either sign, so that the fixed point can be physical for both $B < 0$ and $B > 0$. 

When examining the fixed points of RGEs, an important question is whether FPs can be reached in the UV or IR, and in which particular directions in theory space it is possible. In what follows, we characterize the directions in the coupling space as (IR) relevant if they allow one to reach the fixed point in the UV, and as (IR) irrelevant if they draw couplings away from FP with the increase in the RG scale ({Obviously, the IR-irrelevant directions correspond to the UV relevant ones and vice versa).} Thus, the notion of relevant or irrelevant we employ refers to the orientation of RG flow direction with respect to a particular fixed point. 

If we want to observe how the couplings flow around a given fixed point, we should expand the $\beta$-functions in its vicinity, which leads to the linearised flow for $\delta_i = \alpha_i - \alpha_i^*$
\begin{align}
	\partial_t \delta_i  = \partial_j \beta_i (\alpha^*)  \delta_j + O(\delta^2)
	\equiv 
	- \omega_{ji} \delta_j + O(\delta^2)
	\label{eq:linearized_rge}
\end{align}

The stability matrix $\omega_{ij}$ is given by the first derivatives of beta-functions and is not necessary symmetric. The eigenvalues $\theta_k$ of $\omega$ and the corresponding left eigenvectors $c_i^{(k)}$,
\begin{align}
	c_{i}^{(k)} \omega_{ij} = \omega_k c_{j}^{(k)},
	\label{eq:omega_eigen}
\end{align}
can be used to solve the linearised RGE \eqref{eq:linearized_rge} in the form  
\begin{align}
	(\alpha_i(\mu) - \alpha_*) %
	= 
\sum\limits_{k} c_i^{(k)} \left(\frac{\mu}{\mu_0} \right)^{-\theta_k} 
c_{(k)}^j (\alpha_j(\mu_0)  - \alpha_j^*)
	,
	\label{eq:linerized_rg_flow}
\end{align}
where the flow ``starts'' from scale $\mu_0$, and
we assume that the matrix $c^{(k)}_{i}$ is not degenerate; thus, it can be inverted to give $c_{(k)}^{i}$.  Equation~\eqref{eq:linerized_rg_flow} encapsulates the features of the \emph{powerlaw-like}  flow around a fixed point. The eigenvalues $\theta_k$ play a role of critical exponents of the RG flow, and their sign determines whether the corresponding eigendirections $\delta_i \propto c_i^{(k)}$ drives $\alpha_i$ away from or closer to the fixed point. More explicitly,  the fixed point can only be reached in the UV ($\mu\gg\mu_0$) if at least one of the eigenvalues is positive. On the contrary,\linebreak  for $\mu\ll\mu_0$, the difference $\delta_i$ increases for $\theta_k>0$.  
Thus, if we are interested in the flow towards IR, the eigenvectors associated with positive (negative) eigenvalues correspond to relevant (irrelevant) IR directions.  Finally, eigenvalues may be encountered that vanish exactly. 
 The directions associated with them are called marginal, and do not change the flow near the fixed point at the first order. However, at higher orders, they may bring couplings to the UV fixed point, in which case they are marginally IR-irrelevant, or away from it, when they are marginally (IR) relevant.

In the following, we briefly discuss the phase diagram for weakly coupled gauge--Yukawa theories.  There are four different cases: In addition to the Gaussian fixed point, gauge theories either display none, the Banks--Zaks, gauge--Yukawa, or the Banks--Zaks and gauge--Yukawa fixed points, depending on the values for $B$, $C$, and $C^\prime$. 
For convenience, we summarise here explicit expressions for the stability matrices, the corresponding eigenvalues and left eigenvectors for the BZ FP:
\begin{align}
	\omega_{BZ}  & = -\frac{B^2}{C}
	\begin{pmatrix}
	    1 & - \frac{D}{C} \\
	0 & - \frac{F}{B}
	\end{pmatrix}, 
	\label{eq:BZ_stability_matrix}
	\\
	\theta^-_{BZ} & = - \frac{B^2}{C}, \quad 
	\theta^+_{BZ}  = F \frac{B}{C}, 
	\label{eq:BZ_critical_exponents}
	\\
	c^-_{BZ}  & = [1, 0], \quad 
	c^+_{BZ}  = \left[\frac{D}{C}, 1 + \frac{F}{B}\right] 
	\label{eq:BZ_eigendirections}
\end{align}
and the GY FP:
\vspace{-3pt}
\begin{align}
	\omega_{GY} &  = 
	- \frac{B^2}{C'}
	\begin{pmatrix}
		\frac{C}{C'} & \frac{E}{F} \left( 1 - \frac{C}{C'}\right) \\
		- \frac{F}{E} \frac{F}{B} & \frac{F}{B}
	\end{pmatrix},
	\label{eq:GY_stability_matrix}
	\\
	\theta^\pm_{GY} & = - \frac{B^2}{2 C'}
	\left(
		\frac{F}{B} + \frac{C}{C'} \pm \sqrt{
			\left(
				\frac{F}{B}+ \frac{C}{C'}
			\right)^2
			- 4 \frac{F}{B}
		}
	\right),
	\label{eq:GY_critical_exponents}
	\\
	c^\pm_{GY}  & = \left[
	\frac{E B}{F^2} 
	\left(
		\frac{F}{B} - \frac{C}{C'} \mp 
		\sqrt{
			\left(\frac{F}{B}
			+ \frac{C}{C'}
		\right)^2 - 4 \frac{F}{B}
	}
	\right), 2
	\right].
	\label{eq:GY_eigendirections}
\end{align}

\begin{figure}[H]
\hspace{-7pt}
\begin{tabular}{cc}
	\includegraphics[width=0.45\textwidth]{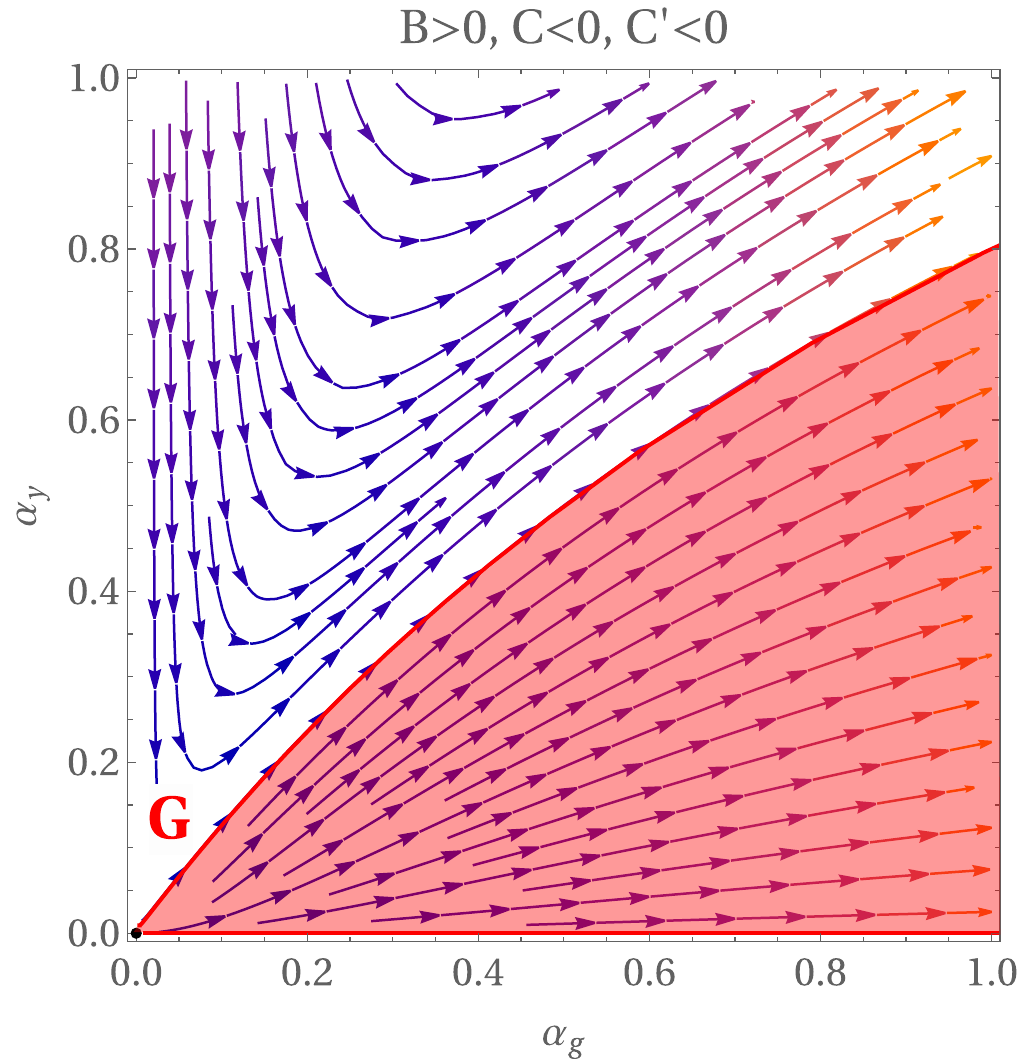} & 
	\includegraphics[width=0.45\textwidth]{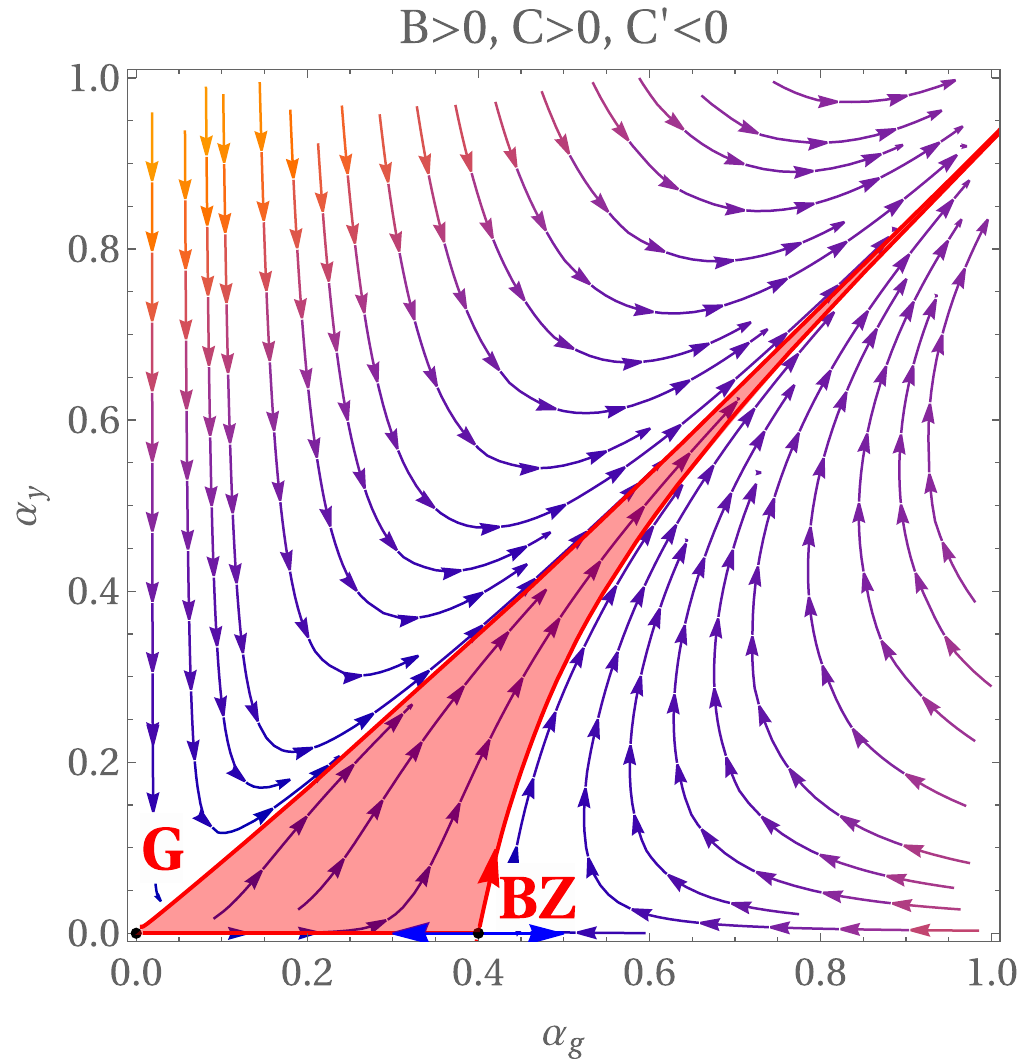}  \\
	\includegraphics[width=0.45\textwidth]{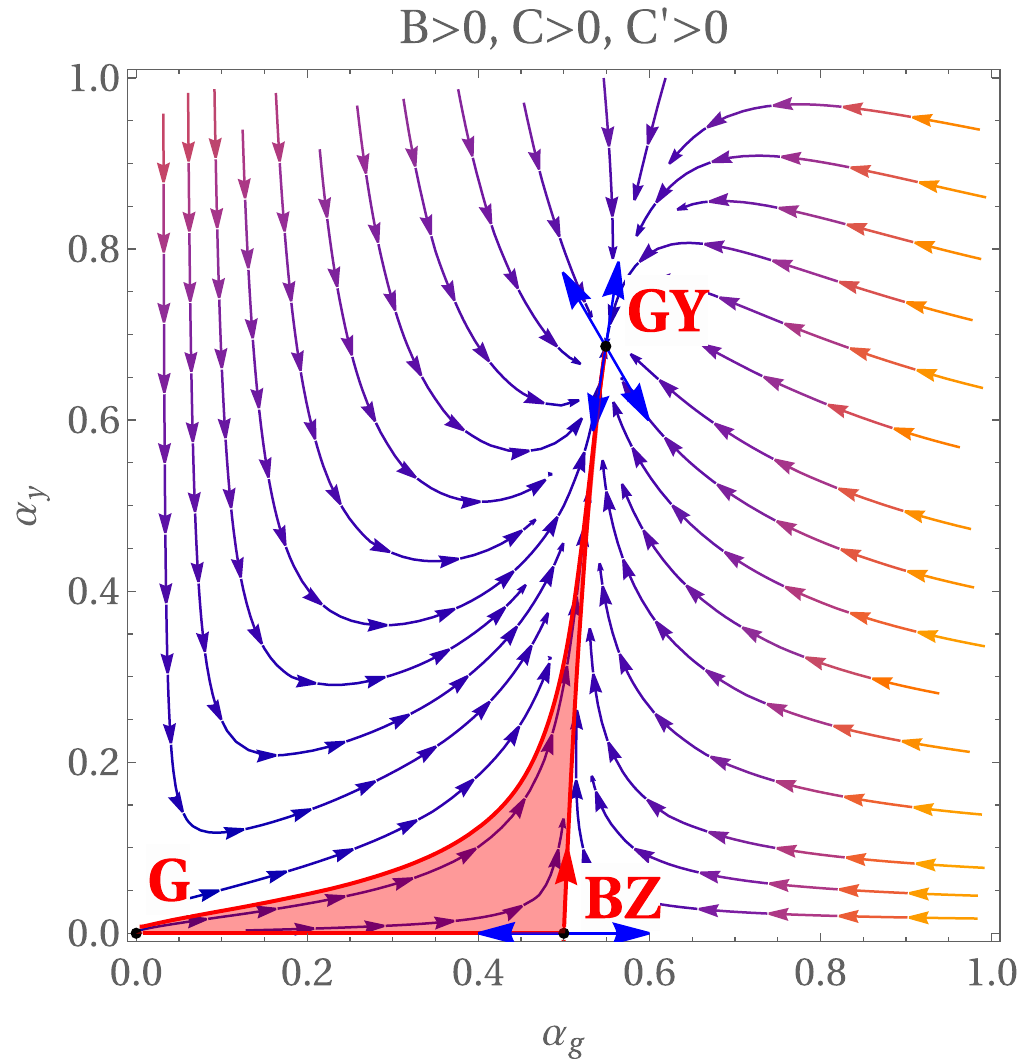} & 
	\includegraphics[width=0.45\textwidth]{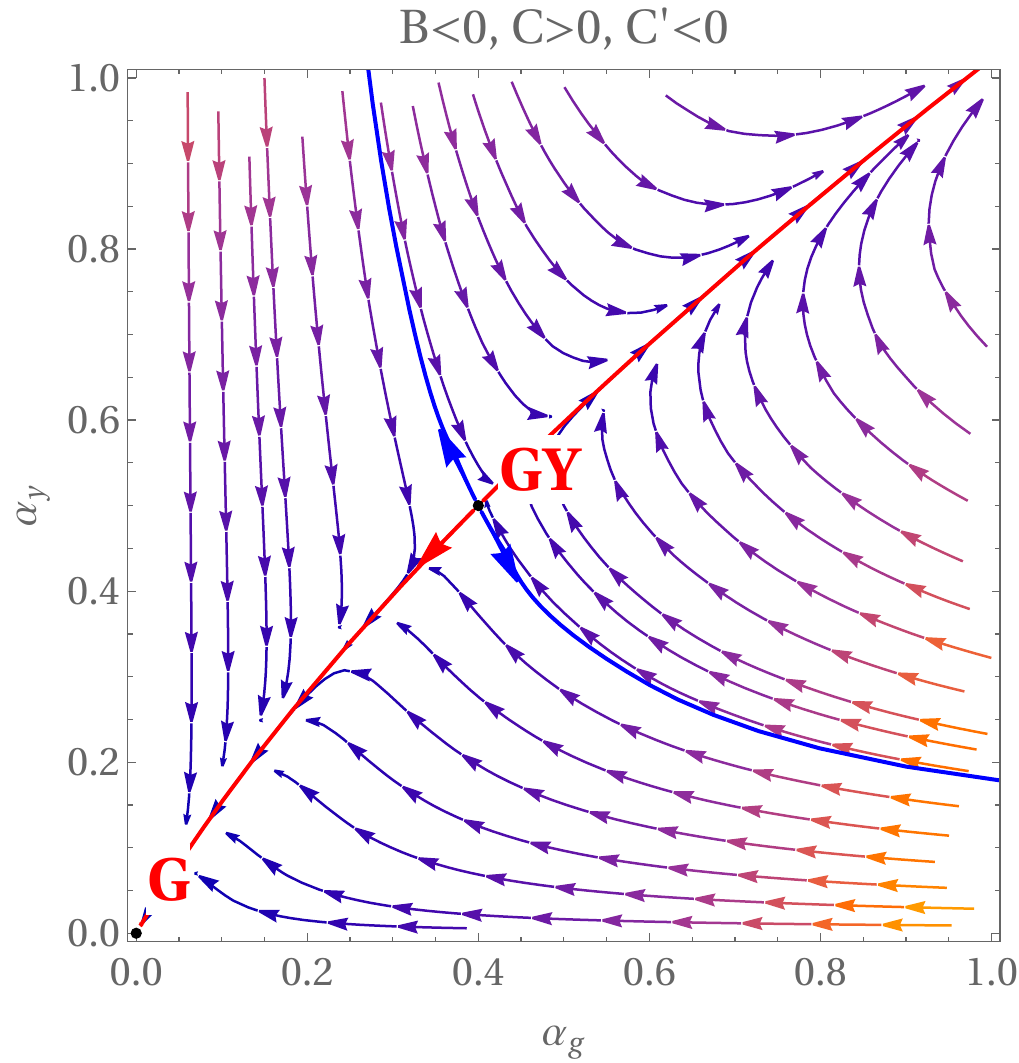}  \\
\end{tabular}
\caption{Phase diagrams of gauge--Yukawa theories. RG flow is towards the IR. Gaussian (G), Banks--Zaks (BZ), and gauge--Yukawa (GY) fixed points are indicated. We also demonstrate IR-relevant (irrelevant) eigendirections for BZ \eqref{eq:BZ_eigendirections} and GY \eqref{eq:GY_eigendirections} FP in red (blue) colour. Shaded areas correspond to UV-complete regions. Adopted from Ref.~\cite{Bond:2016dvk}.}
\label{fig:phase_diagr}
\end{figure}

From the previous discussion, one can observe that BZ FP can exist only for $B>0$.\linebreak  As a consequence,  $\theta^-_{GZ}$ corresponds to the IR-attractive direction, while 
	$\theta^+_{BZ}>0$ and is relevant in IR.  
For the perturbative GY fixed point, $\theta^-_{GY}$ also gives rise to the IR-irrelevant direction, while $\theta^+_{GY}>0$ is IR-relevant only for $B<0$. 
The different phase diagrams are presented qualitatively in Figure~\ref{fig:phase_diagr}, projected onto the (gauge, Yukawa) plane. In the following,\linebreak  we provide short comments to Figure~\ref{fig:phase_diagr} \cite{Bond:2016dvk}:
\begin{itemize}
	\item[(a)] For $B > 0$ and $C < 0$, there is no weakly coupled interacting fixed points. At weak coupling, the phase diagram exhibits only asymptotic freedom and a Gaussian UV FP.  The set of UV free trajectories emerging from it is indicated by the red shaded region. Its upper bound is indicated by the Yukawa nullcline $(E \alpha_y = F \alpha_g)$, which also plays the role of an infrared attractor since below it the sign of $\beta_{y}$ \eqref{eq:gy_beta_yu} is negative and controlled by gauge field fluctuations. 
		UV-free trajectories start near Gaussian FP and continue into the strong coupling region, where the theory is expected to exhibit confinement and chiral symmetry breaking, or perhaps a strongly coupled IR-fixed point. One can observe that no trajectories have been found above the Yukawa nullcline that can reach Gaussian FP in the UV. On such trajectories, the theory technically loses asymptotic freedom. Then, the predictivity is restricted to a finite UV scale, unless there is a strongly coupled UV-fixed point somewhere in this region.    
	\item[(b)] For $B > 0$ and $C > 0 > C^\prime$, the theory additionally develops a Banks--Zaks FP that turns out to be perturbative if $B/C$ is sufficiently small.  The Banks--Zaks fixed points are always weakly IR-attractive $\theta^-_{BZ}<0$ in the gauge and strongly IR-repulsive ($\theta^+_{BZ} > 0$) in the Yukawa direction. The first one is due to ~\eqref{eq:gy_beta_g} and follows from the asymptotic freedom, while the second one is from ~\eqref{eq:gy_beta_yu}. Moreover, near the BZ point (and at weak coupling), the flow is parametrically slower in the gauge direction than in the $y$ direction. As a consequence, BZ FP and the Yukawa nullcline play the role of a strong infrared attractive funnel for all flow trajectories emerging from the Gaussian UV FP. This translates into low-energy relations between the Yukawa and the gauge coupling (at weak coupling), irrespective of their UV initial conditions.
    	\item[(c)] The $B > 0$ and $C > C^\prime > 0$ case gives rise to a fully interacting gauge--Yukawa fixed point in addition to Banks--Zaks FP.  The main new effect in theories with $C^\prime > 0$ as compared to theories with $C^\prime < 0$ is the funneling  
    	of flow trajectories in the IR direction of the attractive Yukawa nullcline stops, terminating at the interacting IR-fixed point \eqref{eq:GY_fp}. Moreover, the GY point is indeed attractive both in the gauge direction and in the Yukawa direction ($\theta^\pm_{GY}<0$). 
	\item[(d)] For $B < 0$ and $C^\prime < 0$ \cite{Litim:2014uca}, we observe that there is no asymptotic freedom, and the Gaussian FP has become an IR-fixed point. The Yukawa interaction has transformed the positive two-loop coefficient $C > 0$ effectively to $C^\prime < 0$, which allows us to create an interacting gauge--Yukawa fixed point \eqref{eq:GY_fp}. This fixed point does show IR -attractive ($\theta^-_{GY}<0$)\linebreak  and repulsive ($\theta^+_{GY}>0$) directions (see blue and red vectors in Figure~\ref{fig:phase_diagr}). The former is a consequence of the IR-attractive nature of the Yukawa nullcline, and the latter is due to the infrared freedom of the gauge coupling. The GY FP in this case can be qualified as an asymptotically safe fixed point, since there are two UV finite trajectories emerging from~it. The trajectory that connects GY FP with the Gaussian one in the infrared remains perturbative at all scales. The RG flow in the opposite direction leads to the strong coupling when perturbative analysis can not be trusted and should be supplemented by other considerations. Away from the Yukawa nullcline, no trajectories are found that can reach the GY FP in the UV. On such trajectories, the theory technically loses fundamental asymptotic safety and can only be considered as an effective description. Nevertheless, it has limited predictability (a relation between couplings in the IR due to attraction to the nullcline).
\end{itemize}
\vspace{-3pt}

\section{A Toy Model towards Asymptotic Safety \label{sec:toy_model}}
The authors of Refs.~\cite{Litim:2014uca,Bond:2017lnq} considered a particular realization of the case with $B<0$ and $C'<0$ and demonstrated that the asymptotic safety of gauge--Yukawa theories can be realised under strict perturbative control in models with singlet scalar, vector-like fermions, and non-Abelian gauge fields.  In this section, we review this setup (Litim--Saninno model) with its features, and further motivate its role in constructing SM extensions.

As the starting point, both of these papers introduce an $SU(N_c)$ gauge theory with $N_F$ generations of vector-like fermions $\psi_i$. 
Since vector-like fermions do not contribute to chiral anomalies, their gauge-group representations can be chosen arbitrarily. In what follows, we assume that $\psi_i$ transform in the fundamental representation under $SU(N_c)$. The spectrum of the model also includes $N_F \times N_F$ complex scalars $S_{ij}$ that are singlets under the $SU(N_c)$ symmetry. The model is described by renormalizable interactions
\begin{align}
    \mathcal{L}_{AS} = \Tr[\bar{\psi} i \hat D \psi] + \Tr[(\partial_\mu S)^\dagger (\partial_\mu S)] - y \Tr[\bar{\psi}_L S\psi_R + h.c.] -V(S),
    \label{eq:Lag_toy}
\end{align}
\textls[-10]{where $\hat D = \gamma^\mu D_\mu$ with $D_\mu$ being covariant derivative, and the traces are over gauge and flavour indices. The scalar potential includes single-trace ($u$) and  double-trace ($v$)~interactions:}
\begin{align}
    V(S)= %
    u \Tr[S^\dagger SS^\dagger S] + v\Tr[(S^\dagger S)]^2.
    \label{eq:pot_toy}
\end{align}

The key feature of the Lagrangian \eqref{eq:Lag_toy} is the presence of the Yukawa $y$ coupling, which is required to arrange the interacting UV fixed points. It should be noted that here we neglect all possible mass terms and trilinear scalar interactions. It is also worth pointing out that the single coupling $y$ in Equation~\eqref{eq:Lag_toy} does not account for the most general form of Yukawa interactions. Indeed, the flavour structure of the model allows one to write
\begin{align}
    y_{ijkl}\bar{\psi}_{Li}S_{jk}\psi_{Rl}
    \label{eq:yuk_toy}
\end{align}
with indices of the tensor coupling $y_{ijkl}$, each taking values $i, j, k, l = 1 \ldots N_F$. However,\linebreak  we can drastically reduce the number of parameters by utilizing flavour symmetries.\linebreak  In the absence of all Yukawa terms, the Lagrangian $\mathcal{L}_{AS} $ respects the following global flavour symmetry 
\begin{align}
    U(N_F)^2_\psi=U(N_F)_{\psi_L}\otimes U(N_F)_{\psi_R}\nonumber\\
    U(N_F)^2_S=U(N_F)_{S_L}\otimes U(N_F)_{S_R},
    \label{eq:symm}
\end{align}
corresponding to independent unitary rotations of $\psi_{L,R}$ under $U(N_F)_{\psi_{L,R}}$, and bi-unitary transformations of matrix scalar fields under $U(N_F)_{S_L}\otimes U(N_F)_{S_R}$. The Yukawa coupling $y$ breaks \eqref{eq:symm} down to $U(N_F)^2$, with $U(N_F)^2_S$ identified with $U(N_F)^2_\psi$. Obviously, the coupling of the form \eqref{eq:yuk_toy} completely destroys the flavour symmetry. As a consequence, we can restrict ourselves to Equation~\eqref{eq:Lag_toy} by demanding that the theory should respect $U(N_F)_\psi^2$.  

The crucial fact that was used in the analysis of the model \eqref{eq:Lag_toy} is that its $\beta$-functions give rise to a gauge--Yukawa FP, with is perturbative in the Veneziano \cite{Veneziano:1976wm} limit. The latter consists of taking $N_F,N_c \to \infty$ simultaneously, while keeping the ratio $N_F/N_c$ fixed.\linebreak  To observe the effect of this approximation on $\beta$-functions, let us rewrite them in terms of a small parameter
\vspace{-3pt}
\begin{align}
    \epsilon=\frac{N_F}{N_c}-\frac{11}{2}.
    \label{eq:eps_limit}
\end{align}

For $\epsilon>0$, the screening due to fermions dominates the antiscreening of the gauge degrees of freedom (resulting in $B<0$), while for $\epsilon < 0$, the opposite happens  ($B>0$). 
Expansion in powers of $\epsilon$ indicate  that the gauge--Yukawa fixed point and its critical exponents stay perturbative as long as $\epsilon$ remains small; see details in Refs.~\cite{Litim:2014uca, Bond:2017lnq}.  Here, it is suffice to note that in theories containing non-Abelian gauge interactions together with fermionic and scalar matter, large-$N$ methods  confirm the viability of ultraviolet gauge--Yukawa fixed points.

	In the Veneziano limit, the fixed-point values are controlled by $\epsilon$ and remain perturbative for $\epsilon \ll 1$. For large $N_c$, AS is achieved in appropriately rescaled couplings 
\begin{align}
    \Tilde{\alpha}_g=\frac{N_c g^2}{(4\pi)^2}, \quad \Tilde{\alpha}_y=\frac{N_c y^2}{(4\pi)^2}.
    \label{eq:LS_couplings}
\end{align}

The beta-functions for \eqref{eq:LS_couplings} in the $210$-scheme have the form \eqref{eq:gy_beta_g} and \eqref{eq:gy_beta_yu} with 
\vspace{-3pt}
\begin{align}
  B & = -\frac{4}{3} \epsilon, 
& C & = 25 + \frac{26}{3} \epsilon, %
 & C' & = - \frac{2(57 - 46 \epsilon - 8 \epsilon^2)}{3(13 + 2 \epsilon)}, \\
  D &=  \frac{1}{2} \left( 11 + 2 \epsilon\right)^2, 
& E & = 13 + 2 \epsilon, &  F &  = 6,   %
	\end{align}
	where we neglected $\mathcal{O}(1/N_c^2)$ terms ({the corrections are studied in Ref.~\cite{Bond:2021tgu}}) in the limit $N_c \to \infty$.
	For $\epsilon>0$, the one- and two-loop gauge contributions to $\beta_{\Tilde{\alpha}_g}$ are positive; thus, the Gaussinan FP is IR-attractive in the gauge direction, and there is no Banks--Zaks fixed point ($B<0, C>0$). However, the model features interacting GY FP \cite{Litim:2014uca}
\begin{align}
	\Tilde{\alpha}_g=\frac{2 \epsilon(13 + 2 \epsilon)}{57-46\epsilon-8\epsilon^2}  = \frac{26}{57} \epsilon + \mathcal{O}(\epsilon^2),\nonumber\\
	\Tilde{\alpha}_y=\frac{12\epsilon}{57-46\epsilon-8\epsilon^2} = \frac{4 \epsilon}{9} + \mathcal{O}(\epsilon^2).
    \label{eq:fp_1406.2337}
\end{align}

To the leading order in $\epsilon$, the critical exponents are given by \cite{Bond:2017tbw}
\begin{align}
	\theta_{GY}^+=\frac{104}{171}\epsilon^2, \qquad  \theta_{GY}^-=-\frac{52}{19}\epsilon,
\end{align}
which corresponds to one IR-repulsive and one IR-attractive direction. The latter fixes the Yukawa coupling at all scales in terms of the gauge coupling (or vice-versa). In other words, the value of one of the couplings in terms of the other is a prediction of the setting.

\begin{figure}[H]
\hspace{-5pt}
		\includegraphics[width=0.6\textwidth]{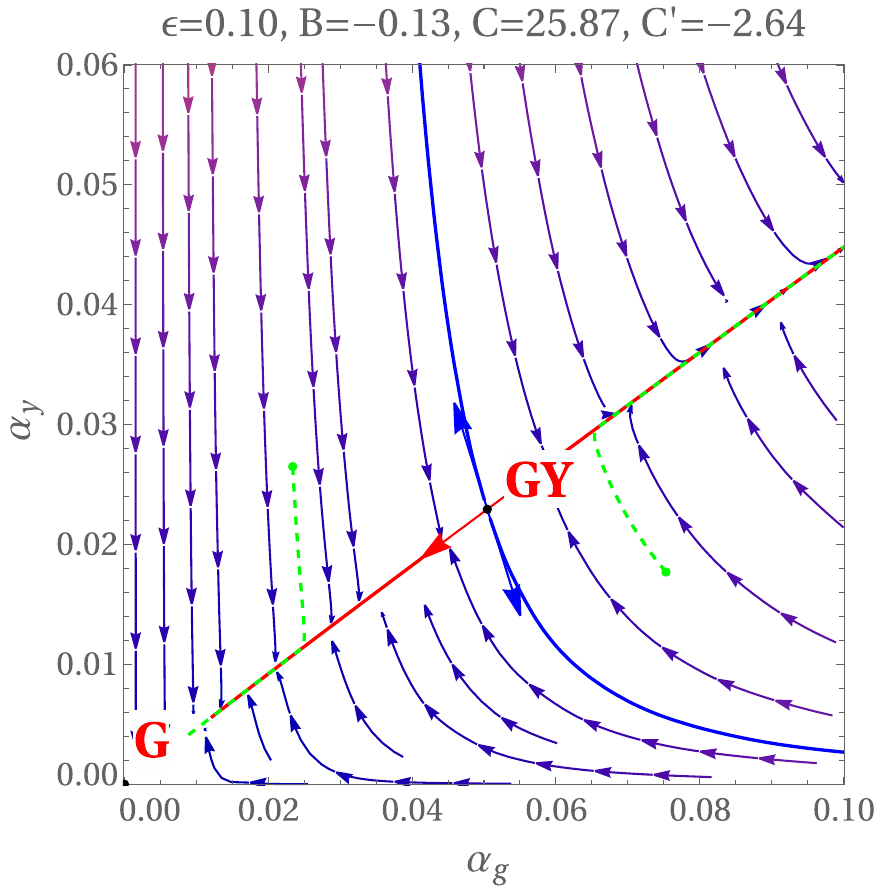}
		\caption{An example of the RG flow for the Litim--Saninno model in the Veneziano limit. The fixed point is given in Equation~\eqref{eq:fp_1406.2337}. The trajectories lying away from the Yukawa nullcline (red line) are rapidly attracted to the latter. For the flow originating below the blue line, the theory remains weakly coupled. In the opposite case, the theory becomes strongly coupled in the IR.}
\label{fig:LS_flow}
\end{figure}

In Figure~\ref{fig:LS_flow}, we demonstrate the flow towards the IR from the fixed point in\linebreak  Equation~\eqref{eq:fp_1406.2337} for a particular value of $\epsilon$. Since $\theta^+_{GY} \sim \epsilon^2$ and $\theta^-_{GY} \sim \epsilon$, for $\epsilon \ll 1$, the flow features one strongly IR-attractive (driven by $\theta^-_{GY}$)  
and one weakly IR-repulsive (corresponding to $\theta^+_{GY}$) direction. 

As discussed earlier, there are two UV-complete (fixed-point) trajectories\linebreak  (red lines in Figure~\ref{fig:LS_flow}) that originate from gauge--Yukawa FP: One ends at the Gaussian FP in the IR, while the other flows to infinity.  Initial conditions in the UV away from the GY FP  result in trajectories that are indistinguishably close to the fixed-point trajectories in the IR. In Figure~\ref{fig:LS_flow}, one can observe two of them (dashed green lines) that start (green dots) below and above the blue curve. The latter separates the regions of weakly and strongly coupled theories in the IR.

In Ref.~\cite{Litim:2014uca}, the authors also studied the model at the next order consistent with WCC ($321$-approximation) and took into account the quartic scalar self-interactions. However,\linebreak  in a more careful study \cite{Bond:2017tbw}, it was argued that instead of a $(n+1,n,n-1)$-approximation, one has to use $(n+1,n,n)$ beta functions to completely determine FPS together with critical exponents up to the order $\mathcal{O}(\epsilon^n)$. In what follows, we consider the $322$ case. For the large-$N_F$-rescaled scalar couplings 
\vspace{-3pt}
\begin{align}
        \Tilde{\alpha}_u=\frac{N_F u}{(4\pi)^2}, \quad \Tilde{\alpha}_v=\frac{N_F^2 v}{(4\pi)^2}
\end{align}
the two-loop beta functions are given by: 
\begin{align}
	\beta^{(1)}_{\Tilde{\alpha}_u} & = -\Tilde{\alpha}_y^2(11+2\epsilon)+4\Tilde{\alpha}_u(\Tilde{\alpha}_y+2\Tilde{\alpha}_u),
	\label{eq:bau_1l} \\
	\beta^{(2)}_{\Tilde{\alpha}_u} & = 
	-24 \Tilde \alpha_u^3 - 16 \Tilde \alpha_u^2 \Tilde \alpha_y 	
	+ 10 \Tilde \alpha_u \Tilde \alpha_g \Tilde \alpha_y
	\nonumber\\
				       & 
	- (11 + 2 \epsilon) 
	\left(
		2 \Tilde \alpha_g \Tilde \alpha_y^2 + 3 \Tilde \alpha_u \Tilde \alpha_y^2
		- (11 + 2 \epsilon) \Tilde \alpha_y^3
	\right),
	\label{eq:bau_2l} \\
	\beta^{(1)}_{\Tilde{\alpha}_v} & = 12\Tilde{\alpha}_u^2 + 4\Tilde{\alpha}_v(\Tilde{\alpha}_v+4\Tilde{\alpha}_u+\Tilde{\alpha}_y), 
	\label{eq:bav_1l} \\
	\beta^{(2)}_{\Tilde{\alpha}_v} & = 
	- 8 \Tilde \alpha_u^2 (12 \Tilde \alpha_u + 5 \Tilde \alpha_v)
	+ 10 \Tilde \alpha_g \Tilde \alpha_v \Tilde \alpha_y
	- 8 ( \Tilde \alpha_u + \Tilde \alpha_v) (3 \Tilde \alpha_u + \Tilde \alpha_v) \Tilde \alpha_y
	\nonumber \\
				       &
	+ (11 + 2 \epsilon)
	\left(
		\Tilde \alpha_y^2 (4 \Tilde \alpha_u - 3 \Tilde \alpha_v) + \Tilde \alpha_y^3
	\right).
	\label{eq:bav_2l}
\end{align}

The gauge beta functions are extended to three loops and that of the Yukawa coupling to two loops, where there is also a contribution due to $\Tilde{\alpha}_u$. One can observe that the double-trace coupling $\Tilde{\alpha}_v$ decouples from the gauge--Yukawa RGE at this order. In the Veneziano limit, the two-loop corrections $\beta^{(2)}_{\Tilde{\alpha}_y}$ to the running Yukawa coupling and the three-loop contributions $\beta^{(3)}_{\Tilde{\alpha}_g}$ to the gauge interaction can be cast into the following form \cite{Litim:2014uca} 
\vspace{-3pt}
\begin{align}
	\frac{\beta^{(2)}_{\Tilde{\alpha}_y}}{\Tilde{\alpha}_y} & =  
          \frac{20\epsilon-93}{6}\Tilde{\alpha}_g^2
	  +(49+8\epsilon)\Tilde{\alpha}_g\Tilde{\alpha}_y 
	  -\frac{11 + 2 \epsilon}{8}
	  \left[
		  (35 + 2 \epsilon) \Tilde{\alpha}_y^2 
		  + 32 \Tilde{\alpha}_y \Tilde{\alpha}_u
	  \right],
	\label{eq:LS_beta_ay_2l} \\
	\frac{\beta^{(3)}_{\Tilde{\alpha}_g}}{\Tilde{\alpha}_g^2} & =  
     \left[ \frac{701}{6}+\frac{53}{3}\epsilon - \frac{112}{27}\epsilon^2\right] \Tilde{\alpha}_g^2 
     + \frac{(11 + 2 \epsilon)^2}{4} 
     \left[
	     (20 + 3 \epsilon) \Tilde{\alpha}_y^2 - \frac{27}{2} \Tilde{\alpha}_g\Tilde{\alpha}_y
     \right]
	\label{eq:LS_beta_ag_3l}
\end{align}

To find FPs as series in $\epsilon$, one can introduce an anzats (in the $322$-approximation) 
\begin{align}
	\alpha_i^* = c^{(1)}_i \epsilon + c^{(2)}_i \epsilon^2  
	\label{eq:FP_anzats_2}
\end{align}
and solve for $c^{(1,2)}_i$. The system of equations $\beta_i(\alpha^*) = 0$ admits a joint, asymptotically safe interacting fixed point with $\Tilde{\alpha}_u>0$, $\Tilde{\alpha}_v<0$, and with $\Tilde{\alpha}_h+\Tilde{\alpha}_v >0$, indicating that at the fixed point, the scalar potential is bounded from below \cite{Litim:2014uca}. The coefficients of \eqref{eq:FP_anzats_2}\linebreak  are given by
  ($X \equiv \sqrt{20 + 6 \sqrt{23}}$)
\vspace{-3pt}
\begin{align}
	& c_g^{(1)}  = +\frac{26}{57}, && c_g^{(2)} =\frac{23 \left(75245-13068 \sqrt{23}\right)}{370386}, %
	\\
	& c_y^{(1)} = + \frac{4}{19}, && c_y^{(2)} = \frac{43549-6900 \sqrt{23}}{20577}, 
	\\
	& c_u^{(1)} = + \frac{1}{19} \left(\sqrt{23}-1\right) , && 
	c_u^{(2)} = \frac{365825 \sqrt{23}-1476577}{631028},
	\\
	& c_v^{(1)}  = - \frac{1}{19} \left(2 \sqrt{23} - X \right), && 
	c_v^{(2)} = -\frac{33533}{6859 X} -\frac{321665}{13718 \sqrt{23}} + \frac{452563}{13718  \sqrt{23} X}+\frac{27248}{6859}
\end{align}
and result in \cite{Bond:2017tbw}%
\vspace{-3pt}
\begin{align}
	\Tilde \alpha_g^* & =  0.45614 \epsilon + 0.780755 \epsilon^2, \\
	\Tilde \alpha_y^* & =  0.210526 \epsilon + 0.508226 \epsilon^2, \\
	\Tilde \alpha_u^* & =  0.199781 \epsilon + 0.440326 \epsilon^2, \\ 
	\Tilde \alpha_v^* & = -0.13725 \epsilon - 0.631784 \epsilon^2.
\end{align}

The corresponding critical exponents can be written as \cite{Bond:2017tbw}
\begin{align}
	\theta_1 & = + \frac{104}{171} \epsilon^2 - \frac{2296}{3249} \epsilon^3 
		 && = &  0.60819 \epsilon^2 - 0.70668 \epsilon^3, \\
	\theta_2 & = -\frac{52}{19} \epsilon + \frac{22783308 \sqrt{23}-136601719}{4094823} \epsilon^2 
		 && = &-2.73684 \epsilon - 6.67594 \epsilon^2,  \\
	\theta_3 & = - X \left[ \frac{8}{19} \epsilon
	- \frac{2 (9153184 \sqrt {23} - 45155739)}{16879999} \epsilon^2\right]
		 && = & -2.94059 \epsilon - 1.04147 \epsilon^2, \\
		\theta_4 & = -\frac{16\sqrt{23}}{19} \epsilon
	+ \frac{4 (255832864 - 68248487 \sqrt{23})}{31393643} \epsilon^2
			 && = & -4.03859 \epsilon - 9.10699 \epsilon^2.
\end{align}

	One can observe that for $\epsilon>0$, the scalar couplings are irrelevant, and again the full model only features one free parameter.

	One important question is related to the range of possible values for $\epsilon$, for which the solution for FP can be trusted (UV conformal window) \cite{Litim:2014uca,Bond:2017tbw}. Limits can arise from the requirements that the theory is weakly coupled $|\Tilde \alpha^*|<1$, 
	the vacuum is stable, and the eigendirection corresponding to the exponent $\theta_1$ remains relevant
 ({vanishing of $\theta_1$ indicates a collision of the UV FP with a non-perturbative IR FP studied in Ref.~\cite{Bond:2021tgu})} ($\theta_1>0)$.\linebreak  A careful analysis carried out in the $322$ approximation, which utilizes the partial information on subleading coefficients, gives rise to \cite{Bond:2017tbw}
	\begin{align}
		0< \epsilon <\epsilon_{max} \approx 0.09...0.13 
		\label{eq:epmax_infNc}
	\end{align}

\begin{figure}[H]
\hspace{-6pt}
	\includegraphics[width=0.7\textwidth]{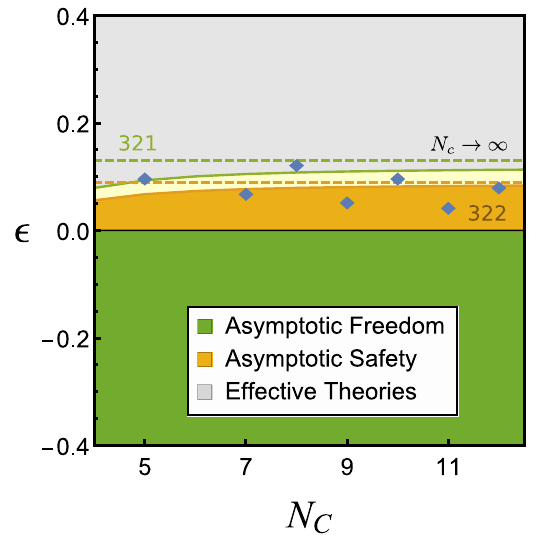}  
	\caption{Conformal window $0<\epsilon<\epsilon_{max}(N_c)$ (from the analysis of subleading terms in beta functions) in the 321 and 322 approximation. Solid curves correspond to $\epsilon_{max}(N_c)$, while the dashed ones
 $\epsilon_{max}(N_c \to \infty)$. Blue dots
 indicate integer values of $N_F$. Adopted from Ref.~\cite{Bond:2021tgu}.
}
\label{fig:conformal_window}
\end{figure}

	Recently, there appeared a study \cite{Bond:2021tgu}, which extends \cite{Bond:2017tbw} and takes into account\linebreak  finite-$N_c$ corrections to the Veneziano limit.  The authors multiplied the expansion coefficients $c_i^{(1,2)}$ \eqref{eq:FP_anzats_2} by functions $f_i^{(1,2)}(N_c)$ that tend to 1 in the limit $N_c\to \infty$, and provided semi-analytical results for these factors. The expressions for critical exponents were also modified appropriately, and $\epsilon_{max}$ was promoted to a function of $N_c$. Based on such corrections, the authors of Ref.~\cite{Bond:2021tgu} concluded that the bound \eqref{eq:epmax_infNc} is lowered for finite $N_c$. Nevertheless, the decrease in the conformal-window size turns out to be moderate\linebreak  (see Figure~\ref{fig:conformal_window}).  

Before switching to more realistic models, let us mention here another limit, which is the large-$N_F$ but finite $N_c$, and formally corresponds to $\epsilon \gg 1$. In this case, matter-field fluctuations dominate and have to be re-summed to all orders \cite{Palanques-Mestre:1983ogz,Gracey:1996he,Holdom:2010qs}.
The studies suggest the existence of an UV Banks--Zaks FP due to a negative singularity of the re-summed beta function. Since this FP may be an artifact of the large-$N_F$ expansion, we do not consider this limit here but refer, e.g., to Refs.~\cite{Antipin:2017ebo,Alanne:2019vuk,Dondi:2019ivp,Kowalska:2017pkt,Antipin:2018zdg,Bond:2021tgu} for more detail and discussion.

\section{SM-like Models with Flavour Portals \label{sec:portals}}

As can be observed, the SM itself has many similarities with the models of Equation~\eqref{eq:Lag_toy}.\linebreak  It is a gauge theory, with Abelian and non-Abelian gauge groups, and it contains fermionic and scalar fields with Yukawa interactions. Therefore, it is natural to wonder whether the SM has ultraviolet fixed points that make it asymptotically safe. However, it is clear that the SM does not exhibit asymptotic safety in the UV, as its $U(1)_Y$ coupling hits a Landau~pole~\cite{Gockeler:1997dn} and the Higgs quartic encounters stability problems (see, e.g., Refs.~\cite{Degrassi:2012ry,Buttazzo:2013uya}). Thus, in~the~SM, the Yukawa couplings are not able to reach a fixed point for all gauge couplings.  Nevertheless,~it may be possible to make it asymptotically safe if the SM is extended.\linebreak  This can be conducted by including new Yukawa interactions that can provide UV FPs for the SM gauge couplings. For example, the new states coupled to the SM through either gauge or Yukawa interactions will eventually modify RGEs of the SM couplings. 
 Thus, minimal extension of the SM and assumptions that the vector-like fermions are charged under the SM gauge group may improve the situation.

Therefore, in the following, we will consider interactions which act as portals between the SM and BSM sector that previously were explored in Ref.~\cite{Bond:2017lnq} and subsequent\linebreak  works~\cite{Barducci:2018ysr,Hiller:2019mou,Hiller:2020fbu,Hiller:2022rla}.  The motivation for this is twofold: Firstly, Yukawa interactions with SM fields is interesting from a phenomenological point of view since it is testable at current experiments. Secondly, the interplay between the SM and BSM fields through Yukawa interactions can provide valuable insights into the flavour sector and its connection to UV completions of the theory. These interactions can play a crucial role in determining the masses and mixing patterns of fermions, such as quarks and leptons, giving rise to observable effects.

Let us start from Ref.~\cite{Barducci:2018ysr}, which explored a large class of models based on the SM matter with $SU(3)_c \times SU(2)_L \times U(1)_Y$ gauge interactions. 
  The authors retain only the top Yukawa coupling together with the Higgs quartic self-interaction and introduce $N_F$ families of vector-like fermions $\psi$ minimally coupled to the SM gauge group and $N_F \times N_F$ generations of scalars $S_{ij}$. These scalars are assumed to be singlets of the SM group.
If we introduce a BSM sector that is charged under $U(1)_Y$, it will cause modifications in the $\beta$-function, allowing us to address the Landau-pole problem that arises in the running of the hypercharge coupling. The Lagrangian characterising this minimal BSM extension is
\begin{align}
    \mathcal{L}_{SM,AS} = \mathcal{L}_{SM} + \mathcal{L}_{AS}.
    \label{eq:lag_SM_AS}
\end{align}

The authors of Ref.~\cite{Barducci:2018ysr} considered 378,000 models with varying numbers of vector-like fermions in different gauge group representations. They conducted a thorough investigation to identify stable, yet perturbative, fixed points within a wide range of parameters corresponding to the number of vector-like fermions and their $SU(3)_c \times SU(2)_L \times U(1)_Y$ quantum numbers. At the end, the authors conclude that the imposed perturbativity conditions are very restrictive. They were not able to find any choice for the group representations and/or number of generations of the vector-like fermions that would make SM reliably asymptotically safe. However, this does not mean that it is definitely not possible to make SM completion asymptotically safe. This implies that if there is an extension of the standard model with AS, it must be different from the models considered by the authors. Otherwise, the fixed point of the model would be beyond the scope of perturbation theory.

Subsequent Refs.~\cite{Hiller:2019mou,Hiller:2020fbu} extended the previous study \cite{Barducci:2018ysr} and considered the role of quartic self-interactions of the scalars $S_{ij}$ as well as \emph{portal} Yukawa and Higgs couplings between SM and BSM. The renormalizable Lagrangian of the models is given by
\begin{align}
    \mathcal{L} = \mathcal{L}_{SM, AS} + \mathcal{L}_{mix} - V(H,S),
    \label{eq:lag_SM_AS_ext}
\end{align}
where the scalar potential fulfills
\begin{align}
    V(H,S) = %
    \delta \Tr[S^\dagger S]H^\dagger H,
    \label{eq:pot_SM_AS}
\end{align}
and $\mathcal{L}_{mix}$ contains Yukawa interactions between BSM and SM matter. 

When speaking about ``portals'', we usually distinguish the following cases. If new particles only couple to the SM gauge fields, we have a ``gauge portal'' (and the models with Lagrangian \eqref{eq:lag_SM_AS} are of this type). One may also introduce new interactions involving the Higgs and the BSM fields such as new Yukawas (``Yukawa portal'') or new quartics\linebreak  (``Higgs portal''). For example, the main effect of gauge portals arises through modifications of the RG-running of the SM interactions due to $\mathcal{L} \subset \bar{\psi}i \hat{D} \psi$, where $\psi$ is again a BSM fermion 
 in a non-trivial representation under the SM gauge group. 
Yukawa portals arise when the Higgs $H$ couples directly to a BSM fermion $\psi$ and a SM fermion $f_{SM}$: $\mathcal{L} \subset \kappa \bar{\psi}H f_{SM}$. The Yukawa portals not only involve new SM charge carriers, but also new interactions controlled by $\kappa$. The new Yukawa coupling  contributes to the running of the Higgs quartic and, thus, influence the vacuum stabilization. 
Finally, Higgs portals arise when the Higgs~$H$~couples to the BSM scalar $S$ through a portal coupling, as in Equation~\eqref{eq:pot_SM_AS}.\linebreak  The inclusion of this new interaction has the advantage of enhancing vacuum stability by contributing positively to the running of the Higgs quartic at a one-loop level.

In Ref.~\cite{Hiller:2020fbu}, six viable models (A--F)  motivated by asymptotic safety were considered. Imposing the condition that at least one Yukawa coupling between the lepton fields $L$, $E$\linebreak  and the vectorlike fermions should be present yields only a few versions for the representations that can take the $\psi$ fields. For example, if $\psi_i$ are singlets with respect to $SU(3)_c\times SU(2)_L$ and have $Y = -1$ hypercharge, their portal interactions $\mathcal{L}_{mix}$ can be written in the form (model A): 
\vspace{-3pt}
\begin{align}
    \mathcal{L}_{mix} = \kappa \bar{L}H\psi_R + \kappa^\prime \bar{E}S^\dagger \psi_L.
\end{align} 

The new Yukawa couplings in $\mathcal{L}_{mix}$ can involve either the SM Higgs or the $S$, and are denoted by $\kappa$ and $\kappa^\prime$ in each case, respectively. The models A--F \eqref{eq:lag_SM_AS_ext} are distinguished solely by the electroweak charges of the vector-like fermions and the allowed Yukawa couplings; see more detail in Refs.~\cite{Hiller:2019mou,Hiller:2020fbu,Hiller:2022rla}.

\subsection*{Portals at Work
 \label{sec:portals_at_work}}
The authors of Refs.~\cite{Hiller:2019mou,Hiller:2020fbu,Hiller:2022rla} found FPs of the $\beta$-functions in the above-mentioned models and explored whether matching to the SM at low energies is possible. They considered constraints from the known values of  $U(1)_Y \times SU(2)_L \times SU(3)_C$ gauge couplings\linebreak  $g_l$ $(l=1,2,3)$, the top and bottom Yukawa interactions $y_{t,b}$, and the Higgs quartic $\lambda$.\linebreak  The SM initial conditions (central values) were applied at the reference scale $\mu_0 = 1$ TeV.

\begin{figure}[H]
\begin{adjustwidth}{-\extralength}{0cm}
\centering
        \includegraphics[width=19cm]{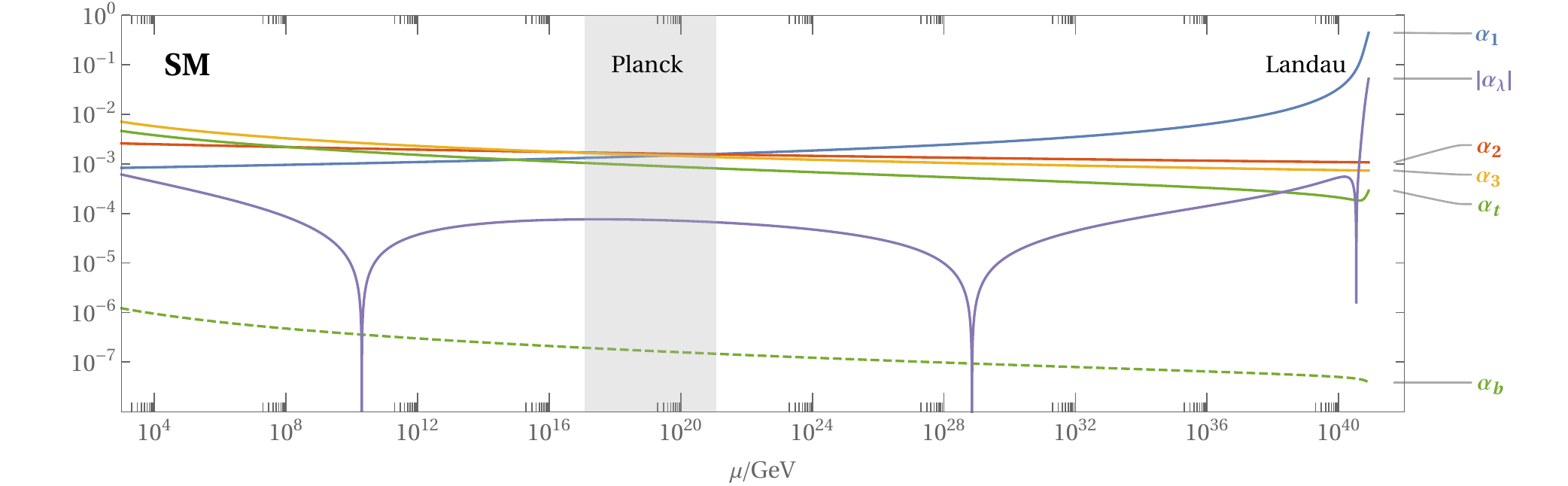}
\end{adjustwidth}
	\caption{The figure shows the SM 3-loop 
		running of the Higgs quartic, top Yukawa, and gauge couplings above TeV energies. The vacuum stability is compromised ($\mu \sim 10^{10}$ GeV) before reaching the Planck scale (center gray band). However, if we disregard quantum gravity effects, the hypercharge coupling can counteract this instability ($\mu \sim 10^{29}$ GeV) and restore stability before perturbativity, stability, and predictivity are ultimately lost at a Landau pole ($\mu \sim 10^{41}$ GeV). Bands indicate a $1 \sigma$ uncertainty in the top pole mass. The picture is taken from Ref.~\cite{Hiller:2020fbu}.}
\label{fig:SM_run}
\end{figure}

From Figure~\ref{fig:SM_run}, we can observe that the SM couplings run slowly.  Refs.~\cite{Hiller:2020fbu,Hiller:2022rla} have integrated the SM RGE from the TeV scale up to the hypercharge Landau pole. The Higgs quartic changes sign $\sim$$10^{10}$~GeV, triggering a well-known vacuum (meta)stability issue. Instead of stopping the flow at this scale, the authors extended it to the trans-Planckian region ignoring quantum gravity effects, and found that the Higgs becomes seemingly stable again $\sim$$10^{10} M_{Pl}$. The vacuum becomes fully unstable at higher scales $\sim$$10^{23} M_{Pl}$. Thus, their conclusion is that additional mechanisms must be introduced to stabilize the vacuum, either at the Planck scale (such as from higher dimensional operators, or full quantum gravity) or below it, e.g., by new particles or interactions.
\vspace{-3pt}

\begin{figure}[H]
\begin{adjustwidth}{-\extralength}{0cm}
\centering
\begin{tabular}{cc}
	\includegraphics[width=0.55\textwidth]{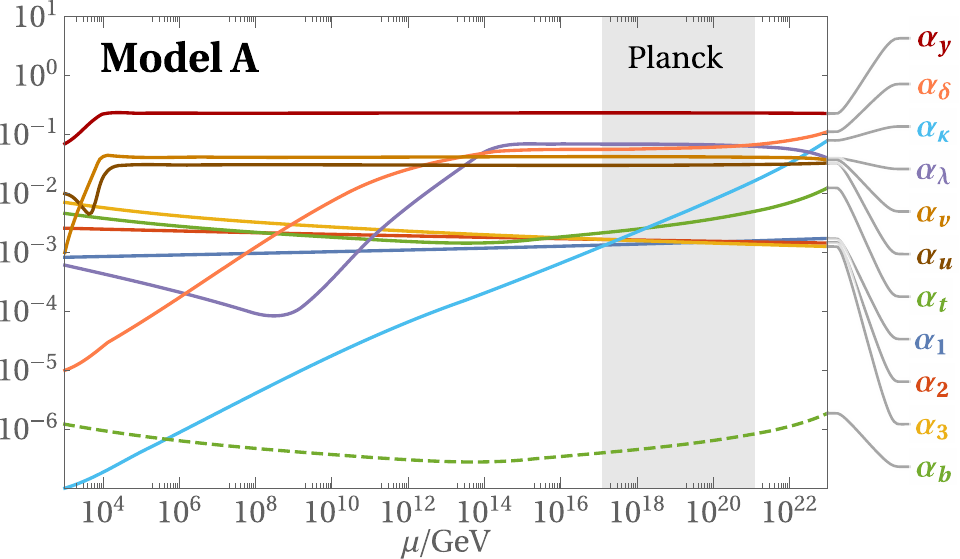} & 
	\includegraphics[width=0.55\textwidth]{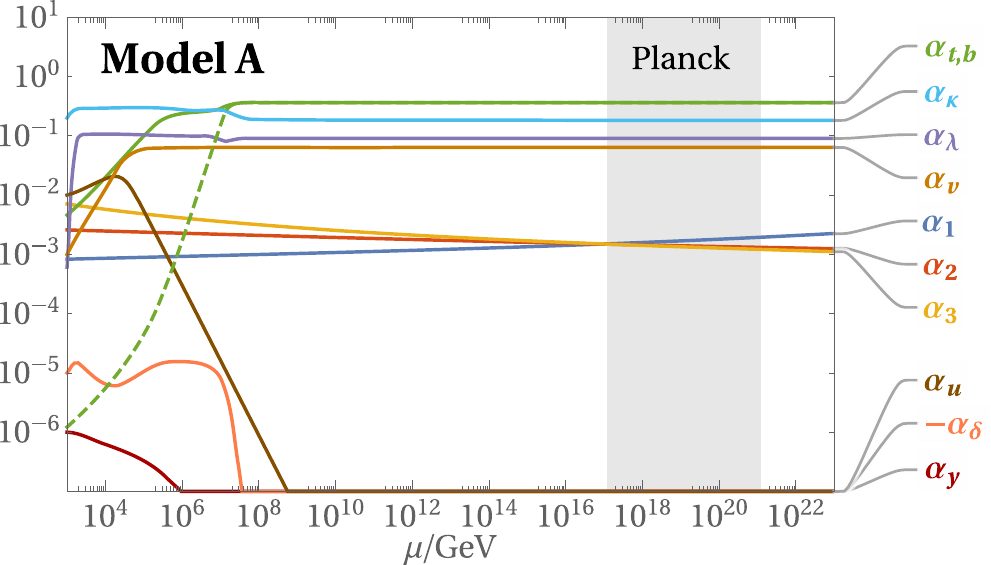} 
\end{tabular}
	\includegraphics[width=1.13\textwidth]{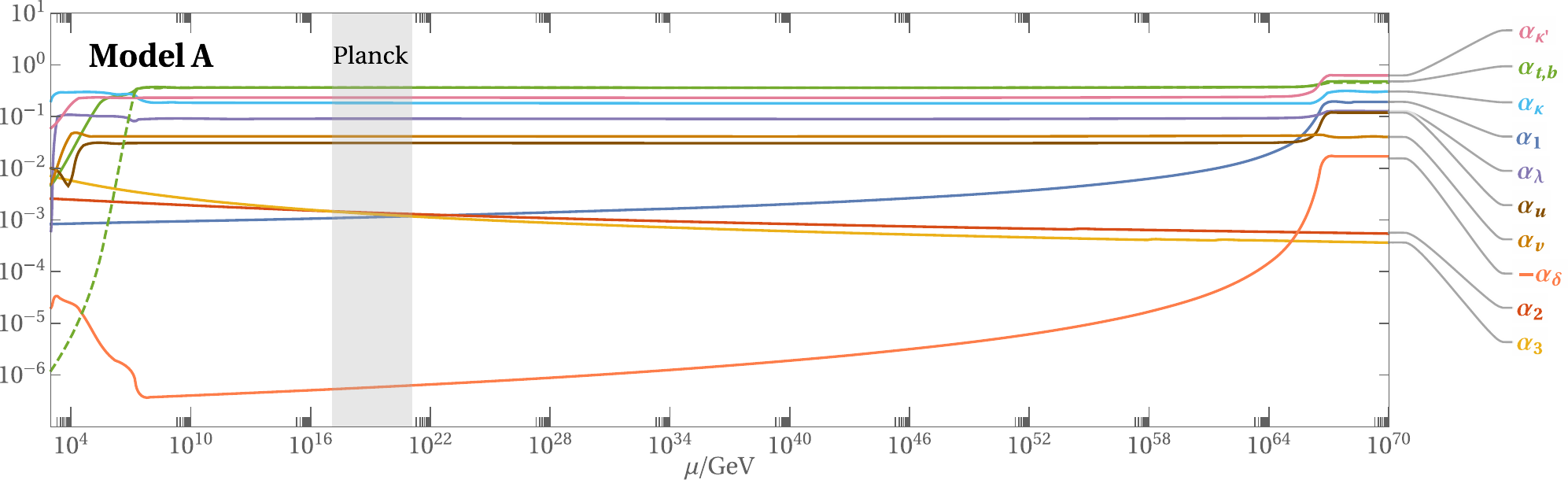} \\ 
	\includegraphics[width=1.13\textwidth]{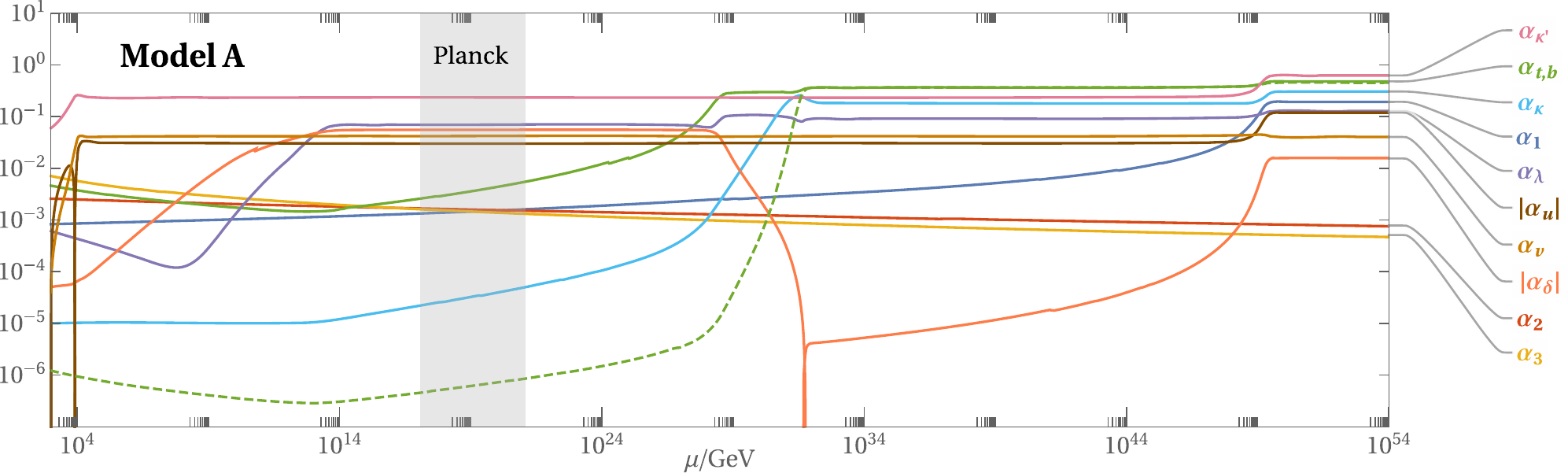}
\end{adjustwidth}  
\caption{Examples of RG flow in model A for different values of BSM couplings at the matching scale. \textbf{Left upper}: $\alpha_y \neq 0$, $\alpha_{\kappa, \kappa^\prime}\sim 0$. \textbf{Upper right}: $\alpha_{y,\kappa^\prime} = 0$, $\alpha_{\kappa}\neq 0$, $|\alpha_{\delta}|\sim 0$. 
\textbf{Middle}: $\alpha_y = 0$, $\alpha_{\kappa,\kappa^\prime} \neq 0$, $|\alpha_{\delta}|\sim 0$.
\textbf{Last}: $\alpha_y = 0$, $\alpha_{\kappa,\kappa^\prime} \neq 0$, $|\alpha_{\delta}|\neq 0$, from 
 Ref.~\cite{Hiller:2020fbu}.}
\label{fig:modelA}
\end{figure}

In the following, we present several examples for model A with a different size for Yukawa couplings $\alpha_y$ at the reference (matching) scale, since  these interactions play a crucial role in avoiding Landau poles and stabilizing RG flows. We indicate scenarios with or without portal coupling $\alpha_\delta$, $\alpha_{\kappa,\kappa^\prime}$ effects in Figure~\ref{fig:modelA}. 

In addition to studying a vacuum-stability issue, the authors of Refs.~\cite{Hiller:2019mou,Hiller:2020fbu,Hiller:2022rla} raised some phenomenological questions, such as the production and decay of BSM particles, fermion mixing, anomalous magnetic moments ($g-2$), effects from scalar mixing, and possible chiral enhancement. They also highlighted signatures at proton--proton and lepton colliders and prospects to detect NP in electric dipole moments or charged lepton-flavour-violating (LFV)-type processes.

\begin{figure}[H]
\hspace{2pt}
        \includegraphics[width=0.75\textwidth]{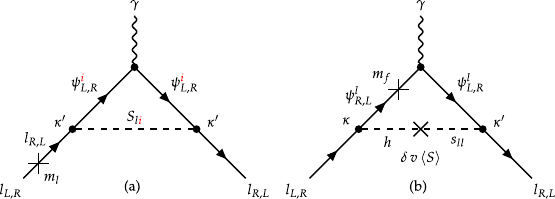}
	\caption{Leading loop contributions to $\Delta a_l$ ($l = e, \mu, \tau$).  (\textbf{a}) BSM scalar--fermion-loops with a lepton chiral flip
(cross on solid line), and (\textbf{b}) chirally enhanced contributions through scalar mixing (cross on dashed line), provided the vacuum exception value $\langle S\rangle\neq 0$, and a BSM fermion $\psi_l$ chiral flip\linebreak  (cross on solid line).}
 \label{fig:amm}
\end{figure}

Let us provide some detail of such phenomenological implications. Ref.~\cite{Hiller:2019mou} considers NP contributions to the muon and electron anomalous magnetic moments. For example, the following two types of Yukawa interactions are introduced
\begin{align}
    \mathcal{L}^{singlet} = -\kappa\bar{L}H\psi_R - \kappa^\prime \bar{E}S^\dagger \psi_L - y\bar{\psi}_L S\psi_R +h.c.,
    \label{eq:Lsing}\\
    \mathcal{L}^{doublet} = -\kappa\bar{E}H^\dagger\psi_L - \kappa^\prime \bar{L}S \psi_R - y\bar{\psi}_L S\psi_R+h.c.,
    \label{eq:Ldoub}
\end{align}
depending on the fact whether $N_F=3$ vectorlike fermions $\psi_{L,R}$ are singlets (corresponding to model A) or doublets under $SU(2)_L$ (model C). The scalar potential is the same as in Equation~\eqref{eq:lag_SM_AS_ext}.
Figure~\ref{fig:amm} demonstrates the relevant leading loop effects due to the new Yukawa $\kappa$, $\kappa^\prime$ and scalar $\delta$ couplings with the additional assumption that due to $S=\langle S \rangle + s$, fermion fields $\psi$ acquire mass $m_f$. Each lepton flavour $l=e,\mu, \tau$ receives a contribution from scalar--fermion loops of the BSM with a chiral flip on the lepton line induced;\linebreak  see Figure~\ref{fig:amm}a). It scales quadratically with the lepton mass \cite{Hiller:2019mou}
\begin{align}
    \Delta a_l = \frac{N_F \kappa^{\prime 2}}{96\pi^2}\frac{m_l^2}{m_f^2}f_1\left(\frac{m_S^2}{m_f^2} \right),
    \label{eq:min}
\end{align}
where $m_S$ is the mass of BSM scalar, and $N_F$ originates from the summation over flavours in the loop in Figure~\ref{fig:amm}a). The function $f_1(t)=(2t^3+3t^2-6t^2\ln t -6t+1)/(t-1)^4$ satisfies $f_1(t)>0$ for any $t\geq 0$; thus, the contribution \eqref{eq:min} is positive and dominant for $a_\mu$.\linebreak  The corrections due to $Z$ and $W$ loops are suppressed parametrically \cite{Hiller:2020fbu}.

If one takes into account Higgs portal coupling $\delta$, there are chirally enhanced\linebreak  contributions, which are linear in the lepton mass (see Figure~\ref{fig:amm}b)). The latter can account for possible deviations in the electron $g-2$ via
\begin{align}
    \Delta a_e=\frac{m_e}{m_f} \frac{\kappa\kappa^\prime \sin 2\theta }{32\pi^2}\left(f_2\left( \frac{m_s^2}{m_f^2}\right)-f_2\left( \frac{m_h^2}{m_f^2}\right) \right)+\frac{m_e^2}{m_\mu^2}\Delta a_\mu,
    \label{eq:chir_enh}
\end{align}
where $m_h$ is the SM Higgs mass. The loop function reads $f_2(t)=(3t^2-2t^2\ln t -4t+1)/\linebreak (1-t)^3$. \textls[-15]{The last term accounts for an additional contribution due to Equation~\eqref{eq:min}. The mixing} \linebreak  angle $\theta$ between the scalar $s_{ll}$ and the physical Higgs $h$ is proportional to $\delta$ \cite{Hiller:2019mou}
\begin{align}
    \tan 2\theta=\frac{\delta}{\sqrt{\lambda(u+v)}}\frac{m_h}{m_s}\left(1+O\frac{m_h^2}{m_s^2} \right).
\end{align}

In summary, the authors concluded that the Yukawa couplings which mix the SM and BSM matter together with a Higgs portal coupling \eqref{eq:Lsing} and \eqref{eq:Ldoub} can generate minimal~\eqref{eq:min} and chirally enhanced \eqref{eq:chir_enh} contributions, which may account for measurements of the muon and electron anomalous magnetic moments. Moreover, as a bonus, they obtained a stable Higgs potential and well-behaved running couplings up to the Planck scale. In addition, a prediction for the deviation of the tau anomalous magnetic moment from its standard model value was provided.

\begin{figure}[H]
\hspace{2pt}
	\includegraphics[width=0.8\textwidth]{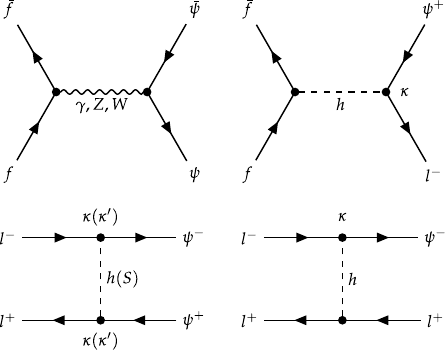}  
\caption{Pair-production of vector-like fermions $\psi$ at $pp$ and $ll$ colliders, with $f$ indicating SM quarks or leptons. Dashed, solid and wavy lines correspond to scalar, fermion, and vector fields, respectively.}
\label{fig:diag_pp_ll}
\end{figure}

In Ref. \cite{Hiller:2020fbu}, the tree-level 
 BSM particle  production at hadron and lepton colliders was discussed in the context of the above-mentioned models. The corresponding diagrams are shown in  Figure~\ref{fig:diag_pp_ll}.

Due to the fact that fermions are assumed to be colourless, the pair production in $pp$ collisions is limited to quark--antiquark fusion to electroweak gauge bosons, as illustrated in the left upper diagram. There is also a possible single production through the Yukawa portal interaction with the $s$-channel Higgs (right upper diagram). In lepton--lepton ($ll$) collisions, $\psi$ can be produced through the $t$-channel Higgs or $S$, either in pairs (as in the left lower diagram) or singly (as in the right lower diagram). 

Eventually, Refs.~\cite{Hiller:2019mou,Hiller:2020fbu,Hiller:2022rla} conclude that the SM extensions with vectorlike fermions are particularly efficient for eliminating the instability of the SM vacuum. This is related to the fact that the gauge portal mechanism enhances the Higgs quartic naturally \cite{Hiller:2020fbu}. 
Further directions towards stability arise in extensions with additional Yukawa/Higgs portals \cite{Hiller:2020fbu,Hiller:2022rla} and anomaly-free gauge interactions \cite{Bause:2021prv}.  Moreover, models with the flavour 
non-diagonal Yukawas or gauge couplings give rise to NP flavour transitions \cite{Hiller:2022rla,Bause:2021prv},  allowing for the alleviation of flavour anomalies. 
 Thus, it would seem interesting to further explore the potential of models inspired by asymptotic safety for flavour and particle physics.
However, despite the many successes of these models, there are still scenarios that suffer from Landau poles in the UV. Therefore, in the next section, we will consider another approach to the AS extension of SM, which takes gravity into account.
\vspace{-3pt}

\section{Models with Gravity and Matter \label{sec:gravity_matter_models}}

The further construction of asymptotically safe models based on SM extensions can be conducted by adding quantum gravity effects. The AS gravity is a powerful way for a Wilsonian description of the fundamental nature of quantum field theories.  In the trans-Planckian regime, it has been proposed \cite{Reuter:1996cp,Reuter:2001ag,Lauscher:2001ya,Manrique:2011jc}  that the quantum fluctuations of the metric field can give rise to an interactive fixed point in the RG flow of the effective action for gravity, which includes the cosmological constant and the Ricci scalar (Einstein--Hilbert truncation). The question related to the persistence of the gravity FP upon the inclusion of  gravitational effective operators of an increasing mass dimension was considered in Refs.~\cite{Lauscher:2002sq,Litim:2003vp,Codello:2006in,Machado:2007ea,Codello:2008vh,Benedetti:2009rx,Dietz:2012ic,Falls:2013bv,Falls:2014tra}, and a positive result was obtained.

The gravity + SM UV fixed point can improve the  high-energy behavior of the hypercharge gauge coupling \cite{Harst:2011zx,Eichhorn:2017lry,Christiansen:2017gtg},  while $SU(3)_c 
\times SU(2)_L$ gauge couplings remain asymptotically free \cite{Daum:2009dn,Daum:2010bc,Folkerts:2011jz}. 

The presence of interacting UV FPs in such a setup may lead to important consequences for its predictivity at low energy, i.e., the actual number of free parameters in the theory can be effectively decreased.  For example,  one can try to predict  the ratio of top and bottom masses \cite{Eichhorn:2018whv}, together with the Cabibbo--Kobayashi--Maskawa \cite{Alkofer:2020vtb}, and Pontecorvo--Maki--Nakagawa--Sakata \cite{Kowalska:2022ypk} matrix elements.

Moreover, AS gravity coupled to the SM demonstrated an early phenomenological achievement by revealing the emergence of an infrared attractive fixed point in the beta function of the Higgs quartic coupling. This finding allowed for a reasonably accurate estimation of the mass of the Higgs boson \cite{Shaposhnikov:2009pv} years prior to its detection at the LHC. As for the recent explorations, there were predictions for the relic abundance of dark matter~\cite{Reichert:2019car, Eichhorn:2020kca}, and analyses of gauged baryon $B$ number \cite{Boos:2022jvc,Boos:2022pyq}, as well as axion models \cite{deBrito:2021akp}.

It is fair to say that it is very hard  to explicitly calculate the quantum-gravity contribution to the matter beta functions of the SM. The pioneering paper by Robinson and Wilzcek \cite{Robinson:2005fj} was criticised by subsequent works (see., e.g., Refs.~\cite{Donoghue:2019clr,Eichhorn:2022gku} and references therein). However, instead of computing 
 these contributions from first principles, some recent studies \cite{Eichhorn:2018whv, Alkofer:2020vtb} have used an efficient approach based on a parametric description of AS gravitational interactions with matter. This phenomenological approach allows one to ``guess'' the strength of the gravitational impact on matter beta  
  functions. The method is based on the assumption that the fixed points of the matter sector should not contradict the low-scale SM phenomenology. The same approach has been used to improve the predictivity of some New Physics  models, for which only incomplete information about their masses and couplings can be obtained experimentally (see, e.g., Refs. \cite{Kwapisz:2019wrl,Domenech:2020yjf,Kowalska:2020gie,Kowalska:2020zve,Kowalska:2022ypk}).

It is generally believed that gravity-induced corrections to matter beta 
 functions are linear in the matter couplings. The phenomenological approach boils down to the following modification of the beta functions of the gauge, Yukawa, and quartic system
\begin{align}
    \beta_g = \beta_g^{SM+NP}-g f_g,\nonumber\\
    \beta_y = \beta_y^{SM+NP}-y f_y, \nonumber\\
    \beta_\lambda = \beta_\lambda^{SM+NP}-\lambda f_\lambda, 
    \label{eq:beta_fg_fy_fl}
\end{align}
i.e., we parameterize the effects of gravitational interactions with effective couplings $f_g$, $f_y$ and $f_\lambda$.  These terms exhibit universality in that gravity does not differentiate between different types of matter interactions (gauge, Yukawa, scalar quartic, etc.), but instead is blind to their internal symmetries.
Note also that in Equation~\eqref{eq:beta_fg_fy_fl}, we disregard any potential quantum gravity effects that are proportional to higher powers in the matter couplings. In the context of complete AS, $f_g$, $f_y$ and $f_\lambda$ should be eventually determined from the gravitational dynamics \cite{Christiansen:2017cxa,Eichhorn:2017eht}.

It should be noted that the aforementioned heuristic approach is based on several simplifying approximations. The parameters $f_g$, $f_y$ and $f_\lambda$ are treated as constants above the arbitrary chosen scale near the Planck mass $M_{Pl} = 10^{19}$ GeV (trans-Planckian region),  and are set to zero below (in the sub-Planckian region). In other words, gravity contributions decouple instantaneously at around  $M_{Pl}$.

\subsection{A Model with Trans-Planckian Asymptotic Safety \label{Sec:AS_grav_work}}

Let us demonstrate the method by applying it to a concrete example \cite{Kowalska:2020zve}. 
As in all previous cases described in Section~\ref{sec:portals}, the authors of Ref.~\cite{Kowalska:2020zve} extended the particle content of the SM by a set of heavy scalar and fermion fields. They add two pairs of fermions and one complex scalar field, belonging to different representations of the $SU(2)_L$ group.\linebreak  The NP Lagrangian can be written in terms of Weyl spinors as
\begin{align}
    \mathcal{L}_{NP} \supset -(Y_R \mu_R E^\prime S + Y_L F^\prime S^\dagger l_\mu + Y_1 E H^\dagger F +  Y_2 F^\prime H E^\prime + h.c. ) - V(H, S), 
    \label{eq:lag_as_grav_model}
\end{align}
where $H$ is the Higgs boson doublet, $l_l = (\nu_{L,l}, e_{L,l})^T$, $E, F$ is two pairs of left-chiral fermion multiplets, and $E^\prime, F^\prime$ is their chiral conjugate. 
 The potential $V(H,S)$ includes quartic self-interactions of $H$ and $S$ and a portal coupling similar to that given in  Equation~\eqref{eq:pot_SM_AS}. 

While the authors of Ref.~\cite{Kowalska:2020zve} consider twelve different charge assignments for the NP fields, we restrict ourselves to the following quantum numbers for new fermions and scalars, charged under the
 $SU(2)_L\times U(1)_Y$:
\begin{align}
    S(\textbf{1},0), \qquad E(\textbf{1},1), \qquad F(\textbf{2}, -1/2).
    \label{eq:M1_quantum_numbers}
\end{align}

Given \eqref{eq:M1_quantum_numbers}, we can derive one-loop beta functions for the hypercharge $g_Y$, strong $g_3$ and weak $g_2$ gauge couplings that have the following form near the Planck scale:
\begin{align}
    \frac{dg_Y}{dt} = \frac{53}{6}\frac{g_Y^3}{16\pi^2} - f_g g_Y,\\
    \frac{dg_2}{dt} = -\frac{5}{2}\frac{g_2^3}{16\pi^2}- f_g g_2,\\
    \frac{dg_3}{dt} = -7\frac{g_3^3}{16\pi^2}- f_g g_3.
\end{align}

To proceed further, one makes the first fundamental assumption: the couplings of the Lagrangian \eqref{eq:lag_as_grav_model} to the gravitational field in the trans-Planckian UV give rise to interactive fixed points.  Furthermore, the fixed-point values associated with the irrelevant directions offer a distinct set of boundary conditions at the Planck scale for the gauge--Yukawa system.

Since we know the measured value of the hypercharge gauge coupling (see,\linebreak  e.g., Refs.~\cite{Buttazzo:2013uya,Bednyakov:2015sca}) at the electroweak scale, it is possible to run it up to the Planck scale with one-loop SM RGE to obtain $g_Y(M_{Pl})$. At the Planck scale, we apply the first fundamental assumption and treat $g_Y(M_{Pl})$ as the fixed-point value:
\begin{align}
    g_Y^* = g_Y(M_{Pl}) = 0.54.
\end{align}

Given $g_Y^*$, we can determine the value of gravity parameter $f_g$ due to the one-loop relation $g_Y^*=4\pi\sqrt{\frac{6f_g}{53}}$:
\begin{align}
    f_g = 0.016.
\end{align}

To agree with the low-energy phenomenology, the non-Abelian gauge couplings are assumed to be asymptotically free:
\begin{align}
    g_2^*=0, \qquad g_3^*=0.
\end{align}

Both $g_2$ and $g_3$ are free parameters of the theory, since they correspond to relevant directions in the couplings space. On the contrary, $g_Y$ corresponds to an irrelevant direction in the coupling space.
It is worth stressing again that the $f_g$ value will be the same for all gauge interactions of the model, since we use the second very important fundamental assumption about the universality of gravity. If that is true, we can immediately read off the FP values of all (additional) gauge couplings. After that, we can run the system down to low energies and read the values of the gauge couplings at the low scale. This demonstrates how asymptotic safety predictions work.

In the same manner, we can find the second quantum gravity parameter, $f_y$.\linebreak  The latter can be fixed if a UV interactive FP point is determined by one of the SM Yukawa couplings, for example, $y_t$. Therefore, from the beta-function zeroes for $y_t$ and $Y_1$ (under the assumption that $Y_2^* = 0$), one can derive:
\begin{align}
    \begin{cases}
   \frac{9}{2}y_t^{*2}-\frac{17}{12}g_Y^{*2}+Y_1^{*2}=16\pi^2f_y,\\
   3y_t^{*2}+\frac{5}{2}Y_1^{*2}-\frac{15}{4}g_Y^{*2}=16\pi^2f_y.
 \end{cases}
\end{align}

Solving this equation with respect to $y_t^{*2}$, we obtain
\begin{align}
    \frac{33}{4}y_t^{*2}+\frac{5}{24}g_Y^{*2}=24\pi^2 f_y,
\end{align}
and after substitution of the FP expression for $g_Y^*$, we determine 
\begin{align}
    y_t^* = 4\pi\frac{\sqrt{-5f_g+318f_y}}{\sqrt{1749}} = 0.41.
\end{align}

Hence, $f_y$ is trivially found if we match the flow of the top Yukawa coupling towards
the experimentally measured top quark mass 
\begin{align}
    f_y=0.006.
\end{align}

For the remaining SM couplings, we have
\vspace{-3pt}
\begin{align}
    y_b^* = 0, \qquad y_\mu^* = 0,
\end{align}
which are associated with relevant directions. In the BSM sector, the authors \cite{Kowalska:2020zve} selected the following fixed point: 
\vspace{-3pt}
\begin{align}
	Y_1^* & = 4\pi\frac{\sqrt{101f_g+106f_y}}{\sqrt{583}} = 0.78,\qquad Y_2^*=0,
    \\ Y_L^*  & =2\pi\frac{\sqrt{-18f_g+53f_y}}{\sqrt{53}}=0.15,\qquad Y_R^*=4\pi\frac{\sqrt{90f_g+53f_y}}{\sqrt{53}}=1.15,
\end{align}
as required for an NP contribution 
 to $\Delta a_\mu$ consistent with the measured value. It should be noted that alternative fixed-point structures can also lead to phenomenological predictions for $\Delta a_\mu$.

\begin{figure}[H]
        \includegraphics[width=0.85\textwidth]{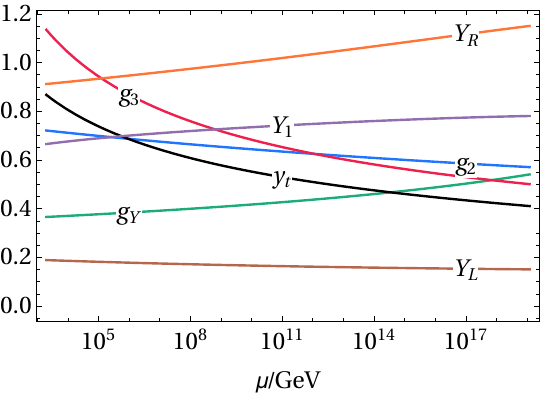}
	\caption{ RG flow of the gauge and Yukawa couplings from the Planck scale down to the reference phenomenological energies of 2 TeV. Above the Planck scale, couplings stabilise and no longer change. The initial values on the Planck scale correspond to the fixed point values.}
\label{fig:coup_run}
\end{figure}

In Figure~\ref{fig:coup_run}, we illustrate the sub-Planckian flow of the parameters of the system for the discussed model.

\subsection{Phenomenological Implications of Trans-Planckian Asymptotic Safety \label{sec:as_gravity_pheno}}

Let us now provide a brief review of some phenomenological implications of the model 
\cite{Kowalska:2020zve}, together with other possible BSM setups \cite{Kowalska:2020gie, Kowalska:2022ypk, Chikkaballi:2022urc}. Here, it should be noted that the following references utilize the Lagrangians in the form \eqref{eq:lag_SM_AS_ext} and \eqref{eq:lag_as_grav_model}; however, with a 
 different kind of particle content. For example, some models contain neutrinos, leptoquarks, or an additional $U(1)'$ gauge $Z^\prime$-boson, etc.

    First of all, Ref.~\cite{Kowalska:2020gie} used an asymptotic safety paradigm to derive predictions for the mass of scalar leptoquarks as solutions to the experimental anomalies noted in recent years in $b\to s$ and $b\to c$ transitions. Using the previously described methods, they found low-energy predictions for the new Yukawa couplings. Then, they combined these predictions with the expectations for the Wilson coefficients in weak EFT extracted from global fits to the full set of $b\to s$ and $b\to c$  transition data. 
After that, they matched those two types of information, and obtained a quite precise determination for the $SU(2)_L$-triplet leptoquark mass at 4-7 TeV from the data on $b\to s$ transitions. These values are too large to be in reach of the high-luminosity LHC. However, according to the most conservative estimates, they are within the early reach of a 100 TeV hadron collider. As for the additional signatures, $BR(K_L\to \mu\mu)$ or $D_0\to \mu\mu$ require significant increases in the experimental sensitivity with respect to the current bounds. 

However, when they applied these methods to the charged-current $b\to c$ anomalies (a different model with a $SU(2)_L$-singlet leptoquark), there arose additional complications due to some tension with low-energy constraints on the fermion masses. Nevertheless, the authors \cite{Kowalska:2020gie} claim that predicted values of mass and Yukawa couplings for the leptoquarks are at the very edge of the current LHC bounds and well within the reach of 300 fb$^{-1}$-integrated luminosity. 
    
Returning back to our example \eqref{eq:M1_quantum_numbers}, the authors~\cite{Kowalska:2020zve} combine the information extracted from the fixed-point UV analysis and bounds from dark matter and collider searches, the measurements of $\Delta (g-2)_\mu$, and the experimental data on $h\to \mu^+\mu^-$ signal strength. These combinations allow them to constrain the favored regions of the parameter space. They found that these results allowed them to pinpoint  the mass of the scalar quite precisely, which reads $m_S\sim$ 100--800 GeV. For other considered models, they obtained the bounds $m_S\sim$ 100--430 GeV or $m_S\sim$ 100--146 GeV. In addition, a strong hierarchy in the fermion spectrum was predicted; the lightest fermion needs to be close in mass to the scalar,  while the mass of the heavier fermion is determined by $\Delta (g-2)_\mu$ and should be around  5--80 TeV in the model considered in this review, and 200--400 GeV or 100--300 GeV for other models. They also found a model with a large region of available parameter space that can be consistent with a TeV-scale dark-matter particle similar to the supersymmetric~higgsino. 

In Ref.~\cite{Kowalska:2022ypk}, the authors considered the SM extended by right-handed neutrinos and investigate the possibility to generate a strong hierarchy in the Yukawa couplings via interplay between the IR-fixed point with zero neutrino Yukawa $y_\nu$ and an UV FP having $y_\nu \neq 0$. They have found the allowed parameter space where Dirac-type neutrino masses can be generated naturally due to the dynamical mechanism. These solutions support the normal mass ordering and are consistent with the current experimental constraints on the mixing parameters.  However, it was stressed that due to the ``blindness'' of gravity, the mixing itself is not a prediction of the fixed-point analysis, as it is associated with relevant directions.    
In addition, a second scenario was considered, in which  sterile right-handed neutrinos constitute a light (sub-MeV) dark matter component of the Universe. In this study, the authors have demonstrated that within the framework of asymptotic safety, the dynamical mechanism naturally produces Yukawa couplings that are consistent with the expected abundance for sterile neutrino dark matter. To achieve additional fixed points in the UV regime, which ensures the completeness of the theory, the introduction of an Abelian gauge interaction and a mirror Yukawa interaction with heavy particles is necessary. In summary, the mechanism proposed in this study offers a UV-complete, generic, and flexible enough solution that can be applied to other models of new physics with feeble Yukawa~interactions.

In Ref.~\cite{Chikkaballi:2022urc}, the authors analyzed two SM extensions with additional $Z^\prime$ boson, vector-like fermions and an SM scalar singlet in the spectrum. Considering the framework of trans-Planckian asymptotic safety, they  provide a solution to the flavour anomalies in the $b\to s\mu\mu$ transitions.
    During the exploration, a fairly precise constraint on the Abelian kinetic mixing $\epsilon$, the NP Yukawa couplings and scalar quartic couplings were derived.  After that, viable mass ranges compatible with $b\to s \mu\mu$ anomalies were extracted and the complete parameter space was subjected to the bounds from the direct production of vectorlike heavy quarks and leptons at the LHC. As a result, the authors identified the parameter space excluded at the 95\% C.L., and computed the projections for the planned increase in luminosity in future runs. 

    As one can oberve, trans-Planckian AS can provide reach phenomenology; we think that the list of possible implications is still far from complete.     

    \subsection{On Robustness of Predictions\label{sec:robustness}}

Recently, there appeared a study \cite{Kotlarski:2023mmr} in which the authors evaluate the precision of the obtained predictions ({by the predictive power of models, the authors mean that given the electroweak values for the SM couplings, it is possible to predict the low-energy values for the NP couplings}) in asymptotically safe gravity-matter models. As it was mentioned earlier, the usual assumptions in such kinds of analyses are the following: (1) The matter beta functions are computed at one loop; (2) the Planck scale is set arbitrarily at $M_{Pl}=10^{19}$ GeV; (3) $f_g$ and $f_y$ are constants above the Planck scale and are zero below the scale. 

The authors drop these assumptions one-by-one and provide estimates of the associated uncertainties. In their exploration, they consider gauged $(B-L)$ and leptoquark SM extensions. This is motivated by the fact that the first type of models has an additional gauge group, for which they seek a prediction for NP gauge couplings. In the second scenario, the key prediction from asymptotic safety is the strength of the NP Yukawa interaction of the scalar leptoquark with the SM fermions.  

To check the robustness of predictions against high-order corrections to beta-functions, which in the case of gauge couplings can be cast in the following form
         \begin{align}
             \partial_t  g_Y = \frac{1}{16\pi^2}(b_Y + \Pi_{n\geq 2}^{(Y)})g_Y^3 - f_g g_Y,
             \label{eq:beta_f_nloop}
         \end{align}
	 where $\Pi_{n\geq 2}^{(Y)}$ collectively denote high-order corrections to one-loop coefficient $b_Y$. The same equations can be written for other gauge couplings. We omit them and discuss only the main idea. Under the assumption that there is an FP at the Planck scale, one derives the $n$-loop expression for $f_g$: 
\vspace{-3pt}
         \begin{align}
			 f_g(n ~ loops) \sim \frac{\left[g_Y^*(n ~ loops)\right]^2}{16\pi^2}(b_Y + \Pi_{n\geq 2}^{(Y)}(g_i^*)). %
         \end{align}
         
	 The authors \cite{Kotlarski:2023mmr} introduce the ratios of the NP gauge couplings $g_i$ and the SM $g_Y$ $r_{g_i}^*\equiv\frac{g_i}{g_Y}$ that do not depend explicitly on the value of $f_g$. This allows one to estimate the uncertainties of low-energy predictions by comparing $r^*_{g_i}$ computed in different loop orders by studying
         \begin{align}
             \frac{\delta r_{g_i}^*}{r_{g_i}^*}=\frac{r_{g_i}^*(2 ~loops)-r_{g_i}^*(1~ loop)}{r_{g_i}^*(1~ loop)}.
         \end{align}
         
	 For simplicity, authors retain only the two-loop corrections and quantify the error at the percent level.
	 A similar but slightly more involved procedure can be conducted for the case of Yukawa couplings. However, in this case, the uncertainty is not negligible, and in the model with scalar leptoquarks can reach tens of the percent \cite{Kotlarski:2023mmr}. 

	 Let us now comment on the arbitrariness related to the position of the Planck scale at which the sub-Planckian RGEs are 
	  matched to trans-Planckian ones. In this respect, the authors consider what happens if gravity decouples from the matter RGEs sharply at a scale that differs from $10^{19}$ GeV by a few orders of magnitude.

When evaluating the effect of the Planck-scale position on the predictions for gauge couplings, one should keep in mind that this uncertainty is effectively equivalent to the uncertainty of the FP value of the hypercharge gauge coupling, $g_Y^*$, and hence $f_g$.\linebreak  From the fact that a ratio of the gauge couplings is considered, it is easy to deduce that the forward-backward moving of the Planck scale does not affect the predicted $r_{g_i}^*$ ratios at the one-loop level, since the dependence of $f_g$ cancels out. In spite of the fact that at higher loops this feature is not preserved, the influence of the Planck scale position remains negligibly small in this case, as well as $O$(0.01\%). 

	 On the contrary, the Yukawa couplings depend explicitly on the fixed-point values of the Abelian gauge couplings, which enter the beta functions. Thus, changing the position of the Planck scale will alter the prediction for the Yukawa couplings even at one loop.\linebreak  The authors \cite{Kotlarski:2023mmr} estimate the uncertainty by considering the ratios of the FP couplings to the reference Yukawa (usually chosen to be that of the top quark) 
         \begin{align}
             \frac{\delta r_{y_i}^*}{r_{y_i}^*}=\frac{r_{y_i}^*(M_{Pl}\neq 10^{19} ~ GeV)-r_{y_i}^*(M_{Pl}= 10^{19} ~ GeV)}{r_{y_i}^*(M_{Pl}= 10^{19} ~ GeV)}.
         \end{align}
         
         Here, the index $i$ in $r_{y_i}$ labels the set of ratios of Yukawa couplings in the Lagrangian. 
         They summarize their findings for two different values of the Planck scale $M_{Pl}=10^{16}$ GeV and $M_{Pl}=10^{20}$ GeV, and conclude that the uncertainties do not exceed $10\%$. 

	 Finally, to address the issue of the potentially scale-dependent $f_{g,y}$, one can study how coupling ratios evolve with scales. At one loop, the gauge coupling ratios turn out to be RG-invariant and, thus, are not affected by the variation of the form of $f_g(t)$. On the contrary, $t$-dependence of $f_{g,y}(t)$ can impact the running of Yukawa ratios starting from one loop.  However, authors conclude that the flow of the ratio $y_i(t)/y_j(t) $ remains fairly stable throughout. Moreover, in this case, the possibility of determining the actual value of the Yukawa couplings at a fixed point is lost. Nevertheless, authors stated that in that range of variability of gravitational parameters, which can be realistically expected in the FRG framework, the obtained uncertainties are moderate.
\section{Conclusions \label{sec:conclusions}}

Asymptotically safe models are of a significant theoretical interest when seeking a comprehensive understanding of fundamental quantum field theories. In our mini-review, we tried to discuss the spectacular progress of the last few years. We started from a general description of the concept and considered simple gauge theories, then we switched to more realistic SM extensions with additional fields, portals, and gravity. A non-exhaustive list of BSM scenarios briefly reviewed here provides just an impression on the asymptotic safety paradigm's prolificity. 

Let us mention that here we do not touch upon many things, including an important subject regarding asymptotic safety in models with  supersymmetry. The latter imposes a relation between the bosonic and fermionic sectors of a theory and can further restrict model building (see, e.g., Refs.~\cite{Intriligator:2015xxa,Bond:2017suy,Abel:2017ujy,Hiller:2022hgt,Bond:2022xvr}, for detail).   

When discussing gravity coupled to matter fields, we intentionally ignore many issues and difficulties reviewed, e.g., in Ref.~\cite{Bonanno:2020bil}. Nevertheless, we are absolutely sure that a new asymptotically safe view on high-energy particle physics seems exciting and potentially useful to explore.

\vspace{6pt}
\authorcontributions{
All the authors contributed equally to all the parts of this work.
 All authors have read and agreed to the text of the manuscript.}

\funding{This research received no external funding.}

\institutionalreview{Not applicable.
}

\informedconsent{Not applicable.}

\dataavailability{Not applicable.
}

\acknowledgments{We thank A.~Baushev, I.~Buchbinder, D.~Fursaev, G.~Kalagov, N.~Lebedev,\linebreak  and I.~Pirozhenko for fruitful discussions.}

\conflictsofinterest{The authors declare no conflict of interest.}

\reftitle{References}

\begin{adjustwidth}{-\extralength}{0cm}

\PublishersNote{}
\end{adjustwidth}

\end{document}